\documentclass[reprint, amsmath, amssymb, aps, prb, superscriptaddress]{revtex4-2}

\usepackage{graphicx}
\usepackage{hyperref}
\usepackage[usenames, dvipsnames]{xcolor}
\usepackage{dcolumn}
\usepackage{bm}
\usepackage{bbold}
\usepackage{enumitem}
\usepackage{titlesec}
\usepackage{orcidlink}
\usepackage{multirow}
\usepackage{xspace}

\usepackage{aligned-overset}

\hypersetup{colorlinks = true, urlcolor = blue, linkcolor = blue, citecolor = blue}

\allowdisplaybreaks

\newcommand{\mb}[1]{\ensuremath{\mathbf{#1}}}
\newcommand{\mc}[1]{\ensuremath{\mathcal{#1}}}
\newcommand{\mr}[1]{\ensuremath{\mathrm{#1}}}

\newcommand{\re}{\ensuremath{\operatorname{Re}}}
\newcommand{\im}{\ensuremath{\operatorname{Im}}}

\newcommand{\one}[0]{\ensuremath{\mathbb{1}}}

\newcommand{\nnnl}{\nonumber \\}

\newcommand{\swave}{{s\!\operatorname{-wave}}}
\newcommand{\Mch}{\ensuremath{\mr{\scriptscriptstyle{M}}}}
\newcommand{\Sch}{\ensuremath{\mr{\scriptscriptstyle{S}}}}
\newcommand{\Tch}{\ensuremath{\mr{\scriptscriptstyle{T}}}}
\newcommand{\Dch}{\ensuremath{\mr{\scriptscriptstyle{D}}}}
\newcommand{\MDch}{\ensuremath{\mr{\scriptscriptstyle{M/D}}}}
\newcommand{\Hart}{\ensuremath{\mr{\scriptscriptstyle{H}}}}
\newcommand{\Mori}{\ensuremath{\mc{\scriptscriptstyle{M}}}}
\newcommand{\overbar}[1]{\mkern 1.5mu\overline{\mkern-1.5mu#1\mkern-1.5mu}\mkern 1.5mu}

\newcommand{\DGA}{D$\Gamma$A\xspace}
\newcommand{\lDGA}{$\ell$D$\Gamma$A\xspace}
\newcommand{\DGAHA}{D$\Gamma$A(HA)\xspace}

\newcommand{\GW}{\ensuremath{{\scriptscriptstyle{GW}}}}
\newcommand{\dga}{\ensuremath{\mr{\scriptscriptstyle{D{\Gamma}A}}}}
\newcommand{\dmft}{\ensuremath{\mr{\scriptscriptstyle{DMFT}}}}
\newcommand{\sbe}{\ensuremath{\mr{\scriptscriptstyle{SBE}}}}

\newcommand{\kN}{\ensuremath{\mathbf{k}_\mathrm{\scriptscriptstyle{N}}}}
\newcommand{\kHS}{\ensuremath{\mathbf{k}_\mathrm{\scriptscriptstyle{HS}}}}
\newcommand{\kAN}{\ensuremath{\mathbf{k}_\mathrm{\scriptscriptstyle{AN}}}}

\newcommand{\Eq}[1]{Eq.~\eqref{#1}}
\newcommand{\Eqs}[1]{Eqs.~\eqref{#1}}
\newcommand{\EQ}[1]{Equation~\eqref{#1}}

\newcommand{\Fig}[1]{Fig.~\ref{#1}}
\newcommand{\Figs}[1]{Figs.~\ref{#1}}

\newcommand{\Sec}[1]{Sec.~\ref{#1}}

\definecolor{betterblue}{RGB}{0,105,175}
\definecolor{darkgreen}{rgb}{0,0.5,0}
\definecolor{darkred}{RGB}{150,0,0}
\definecolor{mygray}{RGB}{150,150,150}

\makeatletter
\def\maketitle{
\@author@finish
\title@column\titleblock@produce
\suppressfloats[t]}
\makeatother

\begin{document}

\title{The finite-difference parquet method: \\ Enhanced electron-paramagnon scattering opens a pseudogap}

\author{Jae-Mo~Lihm\,\orcidlink{0000-0003-0900-0405}}
\thanks{These authors contributed equally to this work}
\email{jaemo.lihm@gmail.com}
\affiliation{European Theoretical Spectroscopy Facility, Institute of Condensed Matter and Nanosciences, Universit\'e catholique de Louvain, Chemin des \'Etoiles 8, B-1348 Louvain-la-Neuve, Belgium}
\affiliation{Department of Physics and Astronomy, Seoul National University, Seoul 08826, Korea}
\affiliation{Center for Theoretical Physics, Seoul National University, Seoul 08826, Korea}

\author{Dominik Kiese\,\orcidlink{0000-0002-9263-8022}}
\thanks{These authors contributed equally to this work}
\email{dkiese@flatironinstitute.org}
\affiliation{Center for Computational Quantum Physics, Flatiron Institute, 162 5th Avenue, New York, NY 10010, USA}

\author{Seung-Sup~B.~Lee\,\orcidlink{0000-0003-0715-5964}}
\email{sslee@snu.ac.kr}
\affiliation{Department of Physics and Astronomy, Seoul National University, Seoul 08826, Korea}
\affiliation{Center for Theoretical Physics, Seoul National University, Seoul 08826, Korea}
\affiliation{Institute for Data Innovation in Science, Seoul National University, Seoul 08826, Korea}

\author{Fabian~B.~Kugler\,\orcidlink{0000-0002-3108-6607}}
\email{fkugler@thp.uni-koeln.de}
\affiliation{Institute for Theoretical Physics, University of Cologne, 50937 Cologne, Germany}
\affiliation{Center for Computational Quantum Physics, Flatiron Institute, 162 5th Avenue, New York, NY 10010, USA}

\date{March 7, 2026}

\begin{abstract} 
We present the finite-difference parquet method that greatly improves the applicability and accuracy of two-particle correlation approaches to interacting electron systems.
This method incorporates the nonperturbative local physics from a reference solution and builds all parquet diagrams while circumventing potentially divergent irreducible vertices.
Its unbiased treatment of different fluctuations is crucial for reproducing the strong-coupling pseudogap in the underdoped Hubbard model, consistent with diagrammatic Monte Carlo calculations.
We reveal a strong-coupling spin-fluctuation mechanism of the pseudogap with decisive vertex corrections that encode the enhanced, energy-dependent scattering amplitude between electrons and antiferromagnetic spin fluctuations.

\bigskip

\textbf{Significance Statement}
Systems of interacting electrons can show emergent, strong-correlation phenomena.
The parquet equations self-consistently relate various propagation amplitudes of many-electron systems and thereby allow one to distill the microscopic mechanisms behind such emergent phenomena.
However, solving them is notoriously hard, particularly when interactions are strong. Here, we present a new method for solving the parquet equations, which circumvents divergences that arise when using nonperturbative input.
We apply our method to the paradigmatic Hubbard model and analyze the pseudogap characterizing the underdoped high-temperature normal state of cuprate superconductors.
We find that an enhanced scattering amplitude between electrons and antiferromagnetic spin fluctuations, or paramagnons, is crucial for forming the strong-coupling pseudogap, and that this enhancement requires the cooperation of multiple paramagnons.

\bigskip

\end{abstract}

\maketitle

Strongly correlated electrons often exhibit complex dynamics. The dynamics can be characterized by the energy and momentum dependence of the electronic correlation functions. Furthermore, their underlying microscopic mechanisms can be revealed by dissecting the correlation functions into vertex functions that describe the effective interactions between various degrees of freedom.

The understanding of correlation and vertex functions in the nonperturbative regime has progressed immensely by computations using the dynamical mean-field theory (DMFT)~\cite{Georges1996} and its nonlocal extensions~\cite{Maier2005,Rohringer2018}. By mapping the lattice problem onto a quantum impurity model, DMFT captures temporal quantum correlations within a given lattice site, which manifest themselves in the strong energy dependence of the local self-energy and vertex functions. Taking a cluster impurity representing multiple lattice sites or a fragmented Brillouin zone, cluster extensions~\cite{Maier2005} can also access short-range spatial correlations. Applied to the Hubbard model (HM) as a proxy of cuprate superconductors, these cluster extensions of DMFT have shown the emergence of a pseudogap (PG) upon increasing the cluster size as well as the importance of spin fluctuations (paramagnons)~\cite{Parcollet2004,Maier2005,Civelli2005,Tremblay2006,Kyung2006, Macridin2006,Haule2007,Ferrero2009,Gull2010,Gunnarsson2015,Wu2018,Gull2013,Meixner2024}.
However, a fine momentum resolution requires large clusters, and thus incurs huge computational costs.

Another route to nonlocal correlations are diagrammatic extensions~\cite{Rohringer2018} of DMFT.
These approaches build long-range correlations on top of the nonperturbative temporal correlations from DMFT calculations.
In other words, they solve field-theoretical relations in which the impurity vertex functions act as renormalized interactions that correlate particles at a distance.
In this way, they naturally reveal which fluctuations and processes are responsible for emergent phenomena.
For example, the dynamical vertex approximation (\DGA)~\cite{Toschi2007,Held2008,Toschi2011,Held2014} solves the parquet equations, which are self-consistent relations between two-particle vertex functions in all three channels---two particle-hole channels and one particle-particle channel.
However, so far, only the ladder \DGA ($\ell$\DGA)~\cite{Rohringer2011,Schaefer2015,Galler2017,Kitatani2020,Kaufmann2021,Schaefer2021, Klett2022,Kitatani2023,Worm2024,Malcolms2024,Bippus2025}---a simplified version that treats a limited subset of two-particle processes---has been utilized to study the PG,
instead of the full parquet \DGA (p\DGA)~\cite{Valli2015,Li2016, Kauch2020, Kauch2019}.
The reason is a hallmark feature of strong correlations: divergences in two-particle irreducible vertex functions~\cite{Schaefer2013,Schaefer2016,Gunnarsson2016,Chalupa2018,Thunstroem2018,Chalupa2021,Reitner2020,Pelz2023,Adler2024}, on which p\DGA is built~\footnote{This problem could be avoided by going from real to dual fermions~\cite{Rubtsov2008, Brener2020}. However, this complicates the interpretation and requires additional approximations as dual fermions have bare interactions between an arbitrary number of particles (instead of the two-particle interaction of real fermions)~\cite{Ribic2017a,Ribic2017b}}.

In this paper, we present a novel formulation of the parquet equations, the finite-difference (fd) parquet method, which is agnostic to singularities in irreducible vertices. We show that access to the \emph{full} vertex of a reference system is sufficient to generate all parquet diagrams of a target system. This allows us to solve the p\DGA of the HM for generic parameters. As an example, we consider a point in phase space markedly close to a vertex divergence, where our p\DGA solution yields a strong-coupling pseudogap consistent with diagrammatic Monte Carlo (DiagMC)~\cite{Wu2017}, while the simplified $\ell$\DGA yields a gapless solution \footnote{For other \lDGA works on the strong-coupling PG, see Refs.~\cite{Klett2022,Worm2024,Malcolms2024,Bippus2025}}.
As a method based on various two-particle vertex functions, fd-p\DGA can be used to distill the microscopic mechanism behind the PG opening.
Our analysis shows that the enhanced electron-paramagnon scattering amplitude (rather than paramagnons themselves) is responsible for the PG opening, and that the diagrammatic origin of this enhancement is the cooperation of antiferromagnetic spin fluctuations in both particle-hole channels.

\section*{Finite-difference parquet method}

The parquet formalism encompasses an exact set of self-consistent quantum many-body relations on the one- and two-particle level~\cite{Bickers2004}. First, the (one-particle) self-energy ($\Sigma$) is related to the two-particle vertex ($F$) by an equation of motion. This Schwinger--Dyson equation (SDE) involves only the full vertex $F$, along with the bare vertex $F_0$, and the full propagator $G$ [see Fig.~\ref{fig:Wu_diagnostics}(a) for a diagrammatic expression \footnote{We ignore the Hartree--Fock part of $\Sigma$ as it can be absorbed in the chemical potential in the present cases.}]. 
In contrast, the self-consistent equations on the two-particle level exploit the parquet decomposition, which categorizes parts of $F$ based on two-particle reducibility. There are three channels, the antiparallel ($a$), parallel ($p$), and transverse antiparallel ($t$) channel, in the parquet equation
\begin{align} \label{eq:F_parquet}
F = \Lambda + \sum_{r=a,p,t} \Phi_r
,
\end{align}

while $\Lambda$ is fully two-particle irreducible (2PI). The $r$-reducible vertex $\Phi_r$ relates to the 2PI vertex $I_r = F - \Phi_r$ in channel $r$ via the Bethe--Salpeter equation (BSE),
\begin{align}
F = I_r + \Phi_r = I_r + I_r \Pi_r F = I_r + F \Pi_r I_r
.
\label{eq:BSE}
\end{align}
Here, the bubble \mbox{$\Pi_r \sim GG$} comprises two full propagators. The multiplication between vertices and bubbles is a generalized matrix operation, see e.g., Refs.~\cite{Kugler2018c, Gievers2022}.

We now consider two systems (reference and target), each defined by a bare propagator and a bare interaction, and analyze the difference of their BSEs.
Further, we denote the ingredients of the BSE for the target (reference) system by the upper-case symbols $\Pi_r$, $F$, $\Phi_r$, $I_r$ (by the lower-case symbols $\pi_r$, $f$, $\phi_r$, $i_r$) and their difference by $\tilde{X} =  X - x$.
From Eq.~\eqref{eq:BSE}, we then derive
\begin{align}
\tilde{\Phi}_r &= I_r \Pi_r F - i_r \pi_r f = \tilde{I}_r \Pi_r F + i_r (\tilde{\Pi}_r F + \pi_r \tilde{F}) 
.
\label{eq:fdBSE_step1}
\end{align}
As the 2PI vertex $i_r$ may be ill-conditioned~\cite{Schaefer2013,Schaefer2016,Gunnarsson2016,Chalupa2018,Thunstroem2018,Chalupa2021,Reitner2020,Pelz2023,Adler2024}, we eliminate it from \Eq{eq:fdBSE_step1}. First, we isolate $\tilde{\Phi}_r$ using $\tilde{F} = \tilde{I}_r + \tilde{\Phi}_r$ on the right-hand side of \Eq{eq:fdBSE_step1},
\begin{align}
\tilde{\Phi}_r & = (\one - i_r \pi_r)^{-1} \bigl[ \tilde{I}_r \Pi_r F + i_r (\tilde{\Pi}_r F + \pi_r \tilde{I}_r) \bigr]
.
\end{align}
Second, we use the BSE in the form 
\mbox{$(\one \!-\! i_r \pi_r)^{-1} i_r \!=\! f$}
and
\mbox{$(\one \!-\! i_r \pi_r)^{-1} \!=\! \one \!+\! f \pi_r$} [cf.~Ref.~\cite{Kugler2018c}, Eqs.~(5,6)] to get
\begin{align}
\tilde{\Phi}_r &= f \tilde{\Pi}_r F + \tilde{I}_r \Pi_r F + f \pi_r \tilde{I}_r \Pi_r F + f \pi_r \tilde{I}_r
.
\label{eq:fdBSE}
\end{align}
Note that, by the left-right symmetry of the BSE, all of the above expressions can alternatively be written with the reference and target functions interchanged.

Suppose we know the solution to the parquet equations for the reference system for a certain choice of the fully 2PI vertex $\Lambda$. We can then find the solution to the parquet equations for the target system for the same choice of $\Lambda$, as $F = f + \sum_r \tilde{\Phi}_r$. By Eq.~\eqref{eq:fdBSE}, this is possible without explicitly invoking $\Lambda$ or any 2PI vertices ($i_r$ or $I_r$). Instead, we only need the full vertices $f$ and $F$ and the \textit{difference} in 2PI vertices, $\tilde{I}_r = \sum_{r' \neq r} \tilde{\Phi}_{r'}$. 
Starting from a well-behaved $f$, $\tilde{\Phi}_r$ obtained from \Eq{eq:fdBSE} through iteration or root finding~\cite{supplemental} remains well-behaved too.

Two notes are in order. First, the case of $\ell$\DGA, where one takes $I_r$ (instead of $\Lambda$) identical between reference and target system, follows immediately from \Eq{eq:fdBSE}: with $\tilde{I}_r = 0$, we have \mbox{$F = f + f \tilde{\Pi}_r F$}
(cf.~Eq.~(27) in Ref.~\cite{Rohringer2018}). Second, \Eq{eq:fdBSE} bears structural similarity to the vertex flow of the multiloop functional renormalization group (mfRG)~\cite{Kugler2018a, Kugler2018b, Kugler2018c}---indeed, its derivation was inspired by the mfRG derivation presented in Ref.~\cite{Kugler2018c}. Here, however, derivatives with respect to the flow parameter are replaced by \emph{finite} differences with respect to a reference solution. In fact, the mfRG vertex flow can be derived from \Eq{eq:fdBSE} by taking the limit of an infinitesimal difference between the reference and the target. In addition, the mfRG vertex flow (where $f \!\to\! F$ and $\pi_r \!\to\! \Pi_r$) is a linear approximation of the nonlinear fd-parquet equations and provides a way to stabilize and accelerate convergence through preconditioning~\cite{supplemental}.

Below, we apply the fd-parquet method to two paradigmatic systems: the Anderson impurity model (AIM) and the Hubbard model (HM). By the $\mr{SU}(2)$ spin symmetry, their BSEs are diagonal in the spin indices after converting the channels $a$, $p$, $t$ to the magnetic ($\mathrm{M}$), density ($\mathrm{D}$), and singlet ($\mathrm{S}$) channels~\cite{supplemental}. For benchmarking, we first consider these models at particle-hole symmetry (PHS). Ultimately, we analyze the PG in the underdoped HM.

\section*{Results}
\subsection*{Anderson impurity model}
For a proof of concept, we start with the AIM at PHS,
\begin{align}
\mathcal{H} & = U \big( n_\uparrow n_\downarrow \!-\! \tfrac{1}{2} n \big)
+ \sum_{\mb{k}, \sigma} \big[ \epsilon_{\mb{k}} c^{\dagger}_{\mb{k} \sigma} c_{\mb{k} \sigma} \!+\! (V_{\mb{k}} d^{\dagger}_{\sigma} c_{\mb{k} \sigma} \!+\! \mathrm{h.c.}) \big]
.
\label{eq:AIM}
\end{align}
Here, $d^{\dagger}_{\sigma}$ ($c^{\dagger}_{\mb{k} \sigma}$) creates a spin-$\sigma$ electron at the impurity (in the bath), $n_{\sigma} \!=\! d^{\dagger}_{\sigma} d_{\sigma}$ and $n \!=\! n_\uparrow + n_\downarrow$. Following Ref.~\cite{Chalupa2021}, we choose $U \!=\! 5.75$, a constant hybridization strength \mbox{$V_{\mb{k}} \!=\! V \!=\! 2$}, and a box-shaped density of states \mbox{$\rho(\epsilon)\!=\! \tfrac{1}{2D}(D - |\epsilon|)$} with half-bandwidth $D \!=\! 10$. 

\begin{figure}
\centering
\includegraphics[width = 0.99\linewidth]{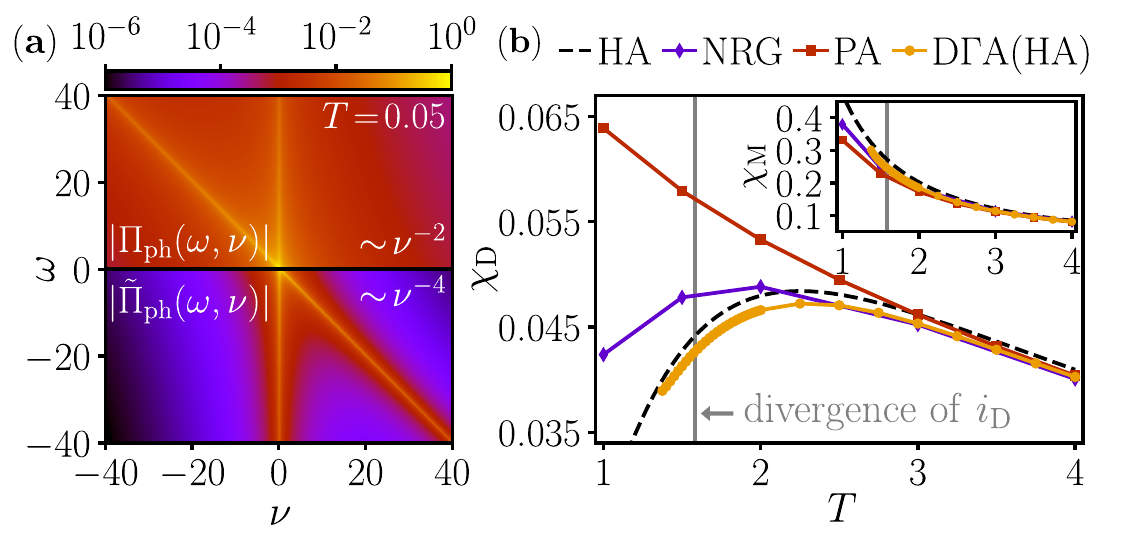}
\caption{(a) Comparison of the two-particle propagator $\Pi$ and its fd counterpart $\tilde{\Pi}$ for an exemplary AIM fd-PA calculation. Both quantities are ($L^{\infty}$-)normalized to reveal the rapid decay of $|\tilde{\Pi}(\omega, \nu)|$ in $\nu$. (b) Charge susceptibility $\chi_{\mathrm{D}}(T)$ in the AIM from different methods. Our fd-\DGAHA employs the Hubbard atom as reference and is soluble irrespective of the 2PI vertex divergence at $T \approx 1.58$. Inset: Spin susceptibility.}
\label{fig:figure_1}
\end{figure}

If \mbox{$\Lambda$} is approximated with the bare vertex $F_0$, as in the parquet approximation (PA), the fd method can be used to accelerate and stabilize the convergence of $F$. 
We illustrate this with the AIM at PHS:
Figure~\ref{fig:figure_1}(a) shows the interacting bubble function obtained in the PA. The fd bubble $\tilde{\Pi}(\omega, \nu)$ decays very rapidly with $\nu$ ($\sim \! \nu^{-4}$ compared to $\sim \! \nu^{-2}$ in $\Pi(\omega,\nu)$) and thus enables a very precise evaluation of the BSE~\cite{vanLoon2024}.
In fact, these results were obtained at a record low temperature $T=0.05$. 
For comparison, in Ref.~\cite{Chalupa2021}, significant numerical resources were required to converge the PA at $T = 0.1$.

Our main focus is combining the fd-parquet equations with non-perturbative solutions of the reference system. For the HM, the prime example is taking the vertex from the DMFT impurity model, so that we obtain a generalization of the \DGA (dubbed fd-\DGA) that circumvents 2PI vertex divergences. Yet, before turning to the HM, we illustrate the effect of a \emph{dynamical} vertex input by taking the Hubbard atom (HA), known analytically~\cite{Thunstroem2018}, as a reference system for the AIM. Starting from $f = F_{\mathrm{HA}}$, \Eq{eq:fdBSE} then incorporates the hybridization between the atom and the bath (cf.~Ref.~\cite{Kinza2013}).
We here denote this approach by \DGAHA.

We find that including the atomic reference considerably improves the plain PA.
As illustrated in Fig.~\ref{fig:figure_1}(b), \DGAHA
captures the local-moment formation in the form of a suppressed charge susceptibility $\chi_{\mathrm{D}}$. 
This non-perturbative feature is absent in the PA~\cite{Chalupa2021}, 
but inherited in \DGAHA from the HA.
Our fd-\DGAHA solution agrees well with benchmark data from the numerical renormalization group (NRG)~\cite{Bulla2008} at $T \gtrsim 2.5$.
At lower $T$, 
the approximation of an atomic $\Lambda$ is insufficient for quantitative agreement and ultimately yields artifacts \footnote{For example, the spin susceptibility becomes negative at some Matsubara frequencies.}.
Importantly, however, \Eq{eq:fdBSE} is soluble even in the presence of the 2PI vertex divergence in $i_{\mathrm{D}}$ at $T \approx 1.58$.

\subsection*{Particle-hole symmetric Hubbard model}
Next, we turn to the square lattice HM,
\begin{align}
\mathcal{H} &= - \sum_{i, j, \sigma} (t_{ij} + \mu \delta_{ij} ) c^{\dagger}_{i \sigma} c_{j \sigma} 
+ U \sum_{i} n_{i \uparrow} n_{i \downarrow} \,,
\end{align}
where $t_{ij} = t$ for nearest neighbors, $t'$ for next-nearest neighbors, and zero else.
Here, $c^{\dagger}_{i \sigma}$ creates a spin-$\sigma$ electron at lattice site $i$, and $n_{i \sigma} = c^\dagger_{i \sigma} c_{i \sigma}$.
We take $t = 1$ as our unit of energy; $U$ denotes the interaction strength.
As the reference solution, we take the propagator $g$ and the full vertex $f$ from the DMFT impurity model solved with the continuous-time interaction expansion quantum Monte Carlo solver (CT-INT)~\cite{Rubtsov2005CTINT} implemented in the TRIQS software library~\cite{Parcollet2015}.
Restarted Anderson acceleration~\cite{Anderson1965,Walker2011} is used to converge vertex and self-energy while simultaneously adjusting the chemical potential to match the DMFT density at finite doping.

Due to the high computational cost associated with solving p\DGA, we approximate the vertices by keeping only the $s$-wave form factor in the truncated-unity expansion of the fermionic momentum dependence of 
$\tilde{\Phi}_r$
\cite{Lichtenstein2017, Eckhardt2018}.
The exact p\DGA is crossing symmetric, but the form-factor truncation slightly breaks this symmetry~\cite{Eckhardt2020}, which is reflected in a shift of the self-energy tail.
We can correct the asymptotic behavior by replacing the local ($\mb{k}$-averaged) p\DGA self-energy with the DMFT one,
\begin{equation} \label{eq:local_corr}
\Sigma_\mb{k}^{\rm p\dga,\, corr}(\nu)
= \Sigma_\mb{k}^{\rm p\dga}(\nu)
+ \Sigma^\dmft(\nu)
- \frac{1}{N_\mb{k}} \sum_{\mb{k}} \Sigma_\mb{k}^{\rm p\dga}(\nu)
.
\end{equation}
This ``local correction'' is motivated by the finding~\cite{Schaefer2021} that the DMFT gives a good account of the local component of the self-energy even if the full self-energy is nonlocal.
Note that the self-energy tail can also be corrected without invoking the DMFT self-energy~\cite{supplemental}.

\begin{figure}
\centering
\includegraphics[width = 0.99\linewidth]{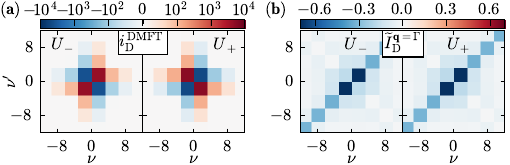}
\includegraphics[width = 0.99\linewidth]{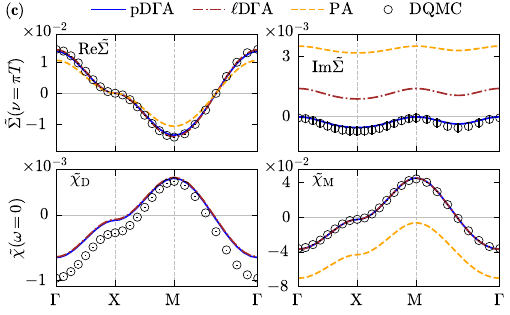}
\caption{
HM at PHS and $U_0= 8.356$, $T = 2$, $t' = 0$.
(a) DMFT 2PI vertex in the density channel 
just below and above the first 2PI vertex divergence 
($U_\pm = U_0 \pm 0.002$)
and (b) its fd counterpart
at $\mb{q}=\Gamma$ obtained from (fd-)p\DGA.
(c) Self-energy and susceptibilities, where the tilde denotes the difference from DMFT, at $U_0$ from (fd-)p\DGA, 
(self-consistent) $\ell$\DGA, PA, DQMC.
The vertical line is the DQMC error bar.
$\tilde{\chi}_\Dch$ in PA is outside the plotted range ($\sim 25 \times 10^{-3}$).
}
\label{fig:highT}
\end{figure}

As a first test, we consider PHS ($\mu = -U/2$, $t' = 0$)
and a high-temperature ($T=2$) and strong-interaction ($U_0 = 8.356$) point in the phase diagram.
This is precisely where the
DMFT 2PI vertex diverges~\cite{Schaefer2013, Schaefer2016}.
Indeed, the inverse of the generalized charge susceptibility (proportional to the 2PI vertex) peaks at a formidable magnitude $O(10^{4})$. Thus, solving the standard p\DGA poses a significant (if not impossible) challenge.

We solve the system via fd-p\DGA at three values of $U$: the point of the 2PI vertex divergence ($U = U_0$) as well as slightly above and below it ($U_\pm = U_0 \pm 0.002$).
The DMFT 2PI vertex $i_{\mathrm{D}}$ has very large magnitude and opposite sign at $U_-$ and $U_+$, see Fig.~\ref{fig:highT}(a).
In contrast, the fd vertex $\tilde{I}_{\mathrm{D}}$ has no singularity, see Fig.~\ref{fig:highT}(b).
In fact, all our results for $U_-$, $U_0$, and $U_+$ are almost on top of each other, demonstrating again that fd-\DGA is robust against crossing 2PI vertex divergences (see \Fig{fig:highT_various_U}~\cite{supplemental}).

We benchmark our fd-\DGA solution with numerically exact determinant quantum Monte Carlo (DQMC) data
\cite{Blankenbecler1981,Assaad2008,Song2025}
in \Fig{fig:highT}(c).
The p\DGA self-energy and susceptibilities agree quantitatively (except for $\chi_\Dch$, which is
more than an order of magnitude 
smaller than $\chi_\Mch$) with DQMC; $\ell$\DGA shows similar agreement, while the PA yields larger deviations.

\begin{figure}[h]
\centering
\includegraphics[width = 0.99\linewidth]{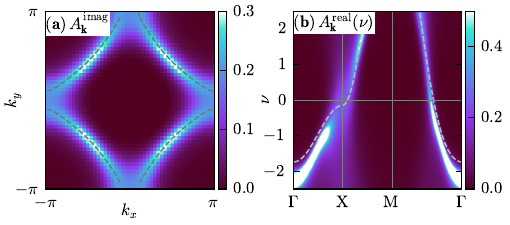}
\caption{
Spectral functions from fd-p\DGA for the HM at $U \!=\! 5.6$, $T \!=\! 0.2$, $t' \!=\! -0.3$, 
and 4\% hole doping.
(a) Matsubara proxy for the spectral function at the Fermi level,
$A_\mb{k}^{\mathrm{imag}} \!=\! -\frac{1}{\pi} \im \frac{1}{\mu - \varepsilon_\mb{k} - \Sigma_\mb{k}(\pi T)}$. 
The gray dashed line denotes the Fermi surface determined by $\mu - \varepsilon_\mb{k} - \re \Sigma_{\mb{k}}(\pi T) \!=\! 0$, which almost coincides with the maximum of $A_\mb{k}^{\mathrm{imag}}$ (cf.\ Ref.~\cite{Simkovic2024}).
(b) Real-frequency spectral function from maximum-entropy analytic continuation~\cite{Kaufmann2023anacont} of the self-energy.
The gray dashed line shows the bare dispersion.
}
\label{fig:Wu_spectral_3d}
\end{figure}

\begin{figure}[t]
\centering
\includegraphics[width = 0.99\linewidth]{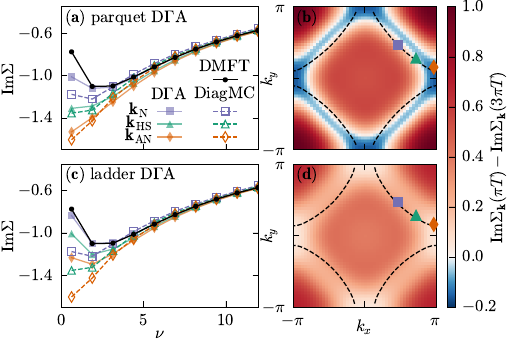}
\caption{
HM at $U = 5.6$, $T = 0.2$, $t' = -0.3$, 
and 4\% hole doping.
(a) Matsubara frequency dependence of $\im \Sigma_\mb{k}(\nu)$
at the node $\kN = (1.47, 1.47)$, hot spot $\kHS = (2.26, 0.88)$, and antinode $\kAN = (3.04, 0.49)$
in p\DGA using \Eq{eq:local_corr}, compared to DiagMC (data from Ref.~\cite{Wu2017}).
(b) Momentum dependence of 
$\Delta \Sigma_\mb{k} = \im \Sigma_\mb{k}(\pi T) - \im \Sigma_\mb{k}(3\pi T)$ in p\DGA.
The dashed line indicates the Fermi surface
obtained from $\varepsilon_\mb{k} + \re\Sigma_\mb{k}(\pi T) = \mu$.
(c,d) Same as (a,b), but for $\ell$\DGA.
}
\label{fig:Wu_self_energy}
\end{figure}

\subsection*{Strong-coupling pseudogap}

A case where strong nonlocal correlations manifest themselves spectacularly is the PG state in the normal phase of doped cuprates.
In angle-resolved photoemission spectroscopy, the PG is characterized by a partial destruction of the Fermi surface, leading to Fermi arcs with reduced weight in the antinodal region near $(0,\pi)$, but not in the nodal region near $(\pi/2,\pi/2)$, of the Brillouin zone~\cite{Ding1996, Marshall1996, Loeser1996}.
This anisotropy and nodal-antinodal dichotomy directly shows the importance of nonlocal correlations.
Further, the importance of spin fluctuations for the PG state was emphasized both in cuprate experiments (e.g., Refs.~\cite{Alloul1989, Nakano1994}) and in HM computations~\cite{Vilk1997, Parcollet2004,Maier2005,Civelli2005,Tremblay2006,Kyung2006,Macridin2006,Haule2007,Ferrero2009,Gull2010,Gunnarsson2015,Wu2017,Wu2018,Gull2013,Schaefer2021,Xu2022,Krien2022a,Yu2024b,Simkovic2024,Sinha2024,Meixner2024}.
While at weak coupling, long-range spin fluctuations induce a PG~\cite{Vilk1997};
at strong coupling, spin fluctuations remain dominant, but their correlation length is limited to a few lattice sites~\cite{Simkovic2024}.

Here, we study the PG in the underdoped HM with parameters for which, as we will demonstrate, p\DGA captures the PG while $\ell$\DGA does not.
Our choice of parameters ($T \!=\! 0.2$, $t' \!=\! -0.3$, $U \!=\! 5.6$, $4\%$ hole doping) follows Ref.~\cite{Wu2017}, where significant numerical resources were used to obtain numerically exact results from diagrammatic Monte Carlo (DiagMC). As shown in Fig.~\ref{fig:Wu_eigenvalues}~\cite{supplemental}, this point is markedly close to a 2PI vertex divergence, illustrating that this PG is in the strong-coupling regime.

To set the scene, we show in Fig.~\ref{fig:Wu_spectral_3d} the spectral function computed with fd-p\DGA.
In Fig.~\ref{fig:Wu_spectral_3d}(a), one can clearly see Fermi arcs, a signature of the PG state, with enhanced (suppressed) spectral weight near the node (antinode).
Figure~\ref{fig:Wu_spectral_3d}(b) further illustrates the reduction of spectral weight around the antinode (close to the X point), which is absent around the node (along the $\Gamma$-M path).
In the following, we systematically distill the microscopic mechanism for this PG. To this end, we trace the PG in the spectral function back to the self-energy, to scattering of collective fluctuations (susceptibilities) with nontrivial amplitudes (Hedin vertices), and ultimately relate these scattering amplitudes to collective fluctuations again.

In diagrammatic approaches, the self-energy can be analyzed very finely in momentum space.
Figure~\ref{fig:Wu_self_energy} compares our p\DGA and $\ell$\DGA self-energies at the node $\kN$, hot spot $\kHS$, and antinode $\kAN$ (see caption for coordinates) to DiagMC~\cite{Wu2017}. 
The p\DGA self-energy is in good agreement with DiagMC [Fig.~\ref{fig:Wu_self_energy}(a)] 
and shows a strong-coupling PG~\cite{Wu2017, Krien2022a, Simkovic2024} [see Fig.~\ref{fig:Wu_self_energy}(a,b)]:
$\Delta\Sigma_{\mb{k}} = \mathrm{Im}\Sigma_{\mb{k}}(\pi T)-\mathrm{Im}\Sigma_{\mb{k}}(3\pi T)$
gradually changes from insulating-like (negative) at $\kAN$ 
through a less insulating $\kHS$ to metallic-like (positive) at $\kN$
\footnote{The local correction for the self-energy tail does not affect the PG: $\Delta\Sigma_{\mb{k}}$ is only weakly affected, and the hierarchy of $\kAN$, $\kHS$, and $\kN$ is unchanged (see \Fig{fig:Wu_tail_compare}~\cite{supplemental}).}.
By contrast, the $\ell$\DGA self-energy shows no PG opening [Fig.~\ref{fig:Wu_self_energy}(c,d)].
We conclude that not only the local input from DMFT but also nonlocal fluctuations of \textit{multiple} channels (p\DGA instead of $\ell$\DGA), 

are required for the strong-coupling PG.

\begin{figure}
\centering
\includegraphics[width = 0.99\linewidth]{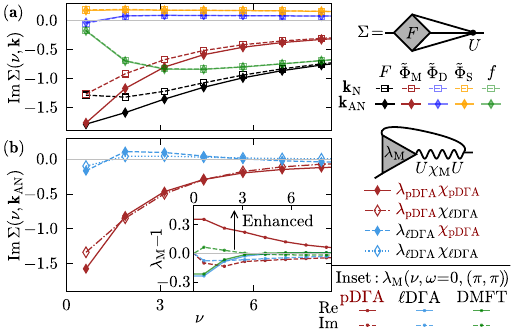}
\caption{
Fluctuation diagnostics (parquet decomposition)~\cite{Gunnarsson2016} for $\Sigma$ from Fig.~\ref{fig:Wu_self_energy}. (a) Result of the SDE at $\kN$ and $\kAN$ from the full vertex $F$, the fd reducible vertices in the magnetic (M), density (D), and singlet (S) channels, and the DMFT impurity vertex ($f$). (b) Interaction-reducible part of the $\tilde{\Phi}_\Mch$ contribution at $\kAN$.
Lines distinguish different combinations of the Hedin vertex and susceptibility from p\DGA and $\ell$\DGA calculations. Inset: Magnetic Hedin vertex at bosonic frequency $\omega=0$ and wavevector $\mb{q} = (\pi, \pi)$ from p\DGA and $\ell$\DGA, showing enhancement and reduction, respectively.
}
\label{fig:Wu_diagnostics}
\end{figure}

To better understand this, we analyze the self-energy by fluctuation diagnostics~\cite{Gunnarsson2015,Schaefer2021a,Dong2022,Yu2024a} (via parquet decomposition~\cite{Gunnarsson2016}).
As shown in Fig.~\ref{fig:Wu_diagnostics}(a), the two dominant contributions to the SDE for $\Sigma$ stem from the local vertex $f$ and from the nonlocal vertex in the magnetic channel $\tilde{\Phi}_{\Mch}$~\cite{Wu2017,Krien2022a,Yu2024b,Simkovic2024}.
While the former contribution is metallic for both the node and antinode, the latter is more strongly insulating at the antinode and thereby drives the PG. Now, if the magnetic channel dominates the nonlocal contributions to $\Sigma$, why is $\ell$\DGA insufficient for the strong-coupling PG?
The reason is that the magnetic Hedin vertex~\cite{Hedin1965, Krien2019b}, i.e., the scattering amplitude between electrons and paramagnons, is non-trivially renormalized, reflecting the cooperation of antiferromagnetic spin fluctuations in both particle-hole channels (here denoted $a$ and $t$, respectively).

Indeed, in Fig.~\ref{fig:Wu_diagnostics}(b), we examine the interaction-reducible part of $\tilde{\Phi}_{\Mch}$ in the SDE \footnote{This part dominates, as can be seen by comparing the red lines in Figs.~\ref{fig:Wu_diagnostics}(a, b).}.
As visualized in the diagram on the right, it can be expressed~\cite{Yu2024a} through the Hedin vertex $\lambda_{\Mch}$ (triangle) and the susceptibility $\chi_{\Mch}$ (wiggly line). Both $\lambda_{\Mch}$ and $\chi_{\Mch}$ are peaked at the antiferromagnetic wavevector $(\pi,\pi)$, signaling paramagnons, and we estimate a relatively short correlation length of $\xi = 5.0$ from $\chi_{\Mch}$~\cite{supplemental}.
Substituting the p\DGA results for either $\lambda_{\Mch}$ or $\chi_{\Mch}$
by the corresponding $\ell$\DGA results, we find that the crucial difference between p\DGA and $\ell$\DGA lies in $\lambda_{\Mch}$.
The antinode's insulating behavior at strong coupling is thus encoded in $\lambda_{\Mch}$.
The inset to Fig.~\ref{fig:Wu_diagnostics}(b) further reveals that the channel cooperation in p\DGA (red) leads to an enhancement of $\lambda_{\Mch}$ ($\mathrm{Re}\, \lambda_{\Mch} \!>\! 1$) at wavevector $(\pi, \pi)$, which is not present in the local DMFT result (green) and is not generated by the single-channel nonlocal dynamics of $\ell$\DGA (blue).
This finding, enabled by our direct access to vertex functions, is our main new result for the strong-coupling PG.
Differently from Ref.~\cite{Krien2022a} but consistently with Ref.~\cite{Simkovic2024}, we did not find a significant influence of 
$\mathrm{Im}\, \lambda_{\Mch}$ on the self-energy
at this $U$ value
\footnote{We note that $\mathrm{Re}\,\lambda_{\Mch}$ in Ref.~\cite{Krien2022a} shows no signs of enhancement.
This might stem from the significant hole doping (15\%) used for the DMFT input.
We also used fd-\DGA to calculate $\Sigma_\mb{k}(\nu)$ for the HM parameters in Fig.~3 of Ref.~\cite{Krien2022a}.
However, our fd-\DGA results were not directly comparable to their dual-fermion calculations, possibly due to large differences in the (hole) doping used for real electrons (15\%) and dual fermions (1\%).}.
Our result is also consistent with recent cluster DMFT studies~\cite{Yu2024a, Yu2024b} finding an enhanced $\mathrm{Re}\,\lambda_{\Mch}$ 
\footnote{Due to the $s$-wave approximation, our calculation cannot resolve the $\mb{k}$-dependence of the vertex function and the Hedin vertex
that was revealed
in Refs.~\cite{Krien2020, Yu2024a, Yu2024b}.}.
The strong $(\omega, \nu)$ dependence of $\lambda_{\Mch}$ (see Figs.~\ref{fig:Wu_Omega_dependence} and \ref{fig:Wu_Hedin_2d}~\cite{supplemental}) supports the conjecture of a strongly energy-dependent vertex, proposed in a DiagMC study \footnote{See the discussion below Eq.~(1) of Ref.~\cite{Simkovic2024}.}.

\begin{figure}[tb]
\centering

\includegraphics[width = 0.99\linewidth]{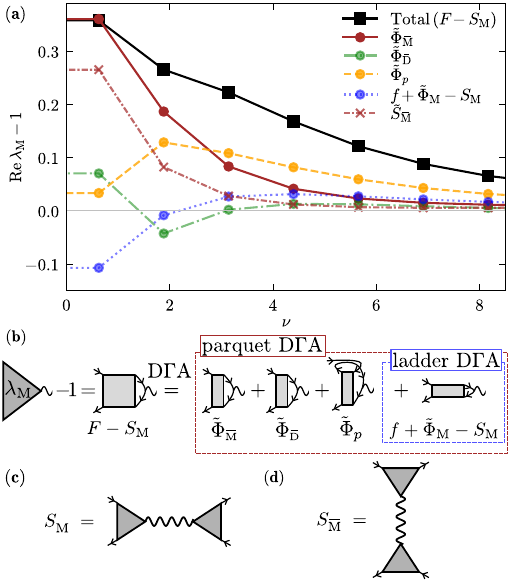}
\caption{
(a) Contributions to the p\DGA magnetic Hedin vertex shown in the inset of \Fig{fig:Wu_diagnostics}(b) according to \Eqs{eq:hedin_bse_simple} and \eqref{eq:hedin_decomp}, represented in (b).
(c, d) Diagrams representing the ($U$-reducible) single-boson exchange magnetic vertices in the (c) `horizontal' ($a$) and (d) `vertical' ($t$) particle-hole channels.
}
\label{fig:Wu_hedin_diagnostics}
\end{figure}

We further analyze the mechanism behind the enhancement of the Hedin vertex using the relation~\cite{Krien2019b}
\begin{equation} \label{eq:hedin_bse_simple}
    \lambda_\Mch
    = 1 - (F_\Mch - S_\Mch) \Pi_t \,,
\end{equation}
where $S_\Mch$ is the single-boson exchange ($U$-reducible) vertex in the magnetic channel [see \Eq{eq:hedin_bse_swave}~\cite{supplemental} for details].
Using the fd decomposition of the vertex, we can write
\begin{equation} \label{eq:hedin_decomp}
    F_\Mch - S_\Mch
    = \tfrac{1}{2} \tilde{\Phi}_{\overbar{\Mch}}
    - \tfrac{1}{2} \tilde{\Phi}_{\overbar{\Dch}}
    + \tilde{\Phi}_{p}
    + \bigl( f_\Mch + \tilde{\Phi}_\Mch - S_\Mch \bigr) \,,
\end{equation}
where the first two terms correspond to the magnetic and density channels in diagrams `transverse' to those of the vertex on the left side (i.e., with shifted frequency arguments)
[see \Eqs{eq:F_fd_spin_basis}--\eqref{eq:transverse_MD_def}~\cite{supplemental}], the third is the $p$ channel contribution, and 
$\tilde{\Phi}_\Mch - S_\Mch$
contains multiboson exchange terms
(2P-reducible but $U$-irreducible)~\cite{Krien2019b, Krien2021}.
Inserting \Eq{eq:hedin_decomp} into \Eq{eq:hedin_bse_simple} yields a decomposition of the Hedin vertex illustrated in \Fig{fig:Wu_hedin_diagnostics}(b), with the $U$-reducible magnetic vertex $S_\Mch$ shown in \Fig{fig:Wu_hedin_diagnostics}(c).

In \lDGA, nonlocal diagrams that mix different diagrammatic channels are neglected,
and only the last term of \Eq{eq:hedin_decomp} is involved in the calculation of $\lambda_\Mch$.
As can be seen from \Fig{fig:Wu_hedin_diagnostics}(a),
this \lDGA contribution (blue circle) is negative, reducing the Hedin vertex.
In p\DGA, on the contrary, all terms are included.
The decomposition of \Eq{eq:hedin_decomp}, shown as circles in \Fig{fig:Wu_hedin_diagnostics}(a), reveals that the Hedin vertex enhancement predominantly comes from the transverse magnetic vertex $\tilde{\Phi}_{\overbar{\Mch}}$ (red circles).
The red crosses in \Fig{fig:Wu_hedin_diagnostics}(a) further show that a sizable contribution comes from $\tilde{S}_{\overbar{\Mch}}$, the $U$-reducible, single-boson exchange part of $\tilde{\Phi}_{\overbar{\Mch}}$,
which is represented in \Fig{fig:Wu_hedin_diagnostics}(d).

\begin{figure}[tb]
\centering
\includegraphics[width = 0.99\linewidth]{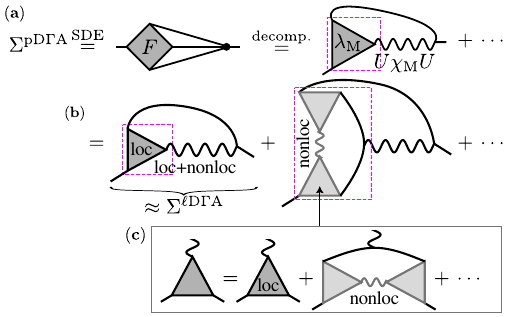}
\caption{
Illustration of the strong-coupling spin-fluctuation mechanism of the pseudogap with vertex corrections (triangles) encoding cooperating spin fluctuations in both particle-hole channels (horizontal and vertical wiggly lines).
(a) SDE for $\Sigma$ and the magnetic single-boson exchange contribution~\cite{Yu2024a}.
(b) The magnetic Hedin vertex (pink dashed boxes) is dominated by a local part and a nonlocal contribution from transverse spin fluctuations $\tilde{S}_{\overbar{\Mch}}$ [indicated by lighter colors; red crosses in \Fig{fig:Wu_hedin_diagnostics}(a)].
(c) This involves Hedin vertices $\lambda_{\overbar{\Mch}}$ which are in turn renormalized by spin fluctuations $\tilde{S}_{\Mch}$, etc.
}
\label{fig:feynman_selfen_vertex}
\end{figure}

In sum, this observation leads to the following qualitative picture of a strong-coupling spin-fluctuation mechanism of the PG.
In \Fig{fig:feynman_selfen_vertex}(a), we adopt the unambiguous fluctuation decomposition~\cite{Yu2024a} and write the self-energy in Hedin form involving the magnetic Hedin vertex and magnetic susceptibility, together with additional terms from the other channels and the $U$-irreducible vertex (not shown).
We focus on the former, as it drives the PG opening [\Fig{fig:Wu_diagnostics}(a)].
Note that we arbitrarily chose a horizontal wiggly line in \Fig{fig:feynman_selfen_vertex}(a).
We could also have chosen a vertical wiggly line---the important point is the interplay between both particle-hole channels.
Now, both the Hedin vertex and the susceptibility contain local contributions from DMFT and nonlocal corrections, as illustrated in \Fig{fig:feynman_selfen_vertex}(b).
To avoid diagrammatic overcounting, a Hedin vertex is by definition irreducible in the bosonic propagator corresponding to the attached wiggly line.
Therefore, in \lDGA, the only nonlocal contributions to $\lambda_\Mch$ come from particle-hole multiboson terms. For the parameters of \Fig{fig:Wu_self_energy}, these terms
are small [see the inset of \Fig{fig:Wu_diagnostics}(b)] and 

hardly affect
the self-energy (see \Fig{fig:Wu_PA_and_trilex}(a)~\cite{supplemental}).
Thus, the \lDGA self-energy captures only the first term in \Fig{fig:feynman_selfen_vertex}(b).

In p\DGA, on the contrary, the magnetic Hedin vertex $\lambda_\Mch$ is renormalized by spin fluctuations in the transverse particle-hole channel, $\tilde{S}_{\overbar{\Mch}}$ [\Fig{fig:Wu_hedin_diagnostics}(d)].
This nonlocal renormalization affects the self-energy through the second diagram in \Fig{fig:feynman_selfen_vertex}(b).
Further, the $\tilde{S}_{\overbar{\Mch}}$ contribution contains Hedin vertices $\lambda_{\overbar{\Mch}}$, which are in turn renormalized by $\tilde{S}_\Mch$ [see \Fig{fig:feynman_selfen_vertex}(c)].
Thereby, the self-consistent feedback between both particle-hole channels with a non-perturbative local seed yields an energy-dependent enhancement of the magnetic Hedin vertex.
This qualitative picture of the strong-coupling PG, with a consistent renormalization of the spin-fluctuation propagator \textit{and} electron-paramagnon scattering amplitude, is another major result of our work.
Yet, we emphasize that all terms in \Eq{eq:hedin_decomp} are needed to numerically reproduce the enhanced Hedin vertex and the strong-coupling PG.

\section*{Discussion and Conclusion}
We developed the fd-parquet method to generate all parquet diagrams of a target system on top of the \textit{full} vertex of a reference system. This allows one to solve the p\DGA while circumventing any singularities of irreducible vertices. We demonstrated the benefits of our approach with calculations of the AIMs at low $T$ and with HA input, of the HM at PHS right at a vertex divergence, and of the doped HM in the PG regime close to a vertex divergence.

The last case is particularly interesting because of its strongly correlated nature (proximate vertex divergence) despite an intermediate strength of $U$ and a relatively high $T$.
It is very different from the PG at weak $U$ and low $T$~\cite{Vilk1997}, where an electron scatters off a long-range paramagnon (that prevails due to a proximate antiferromagnetic instability).
Our case is far away from magnetic order, so the paramagnons are short-ranged and far less pronounced.
Here, the PG opens due to an enhanced electron-paramagnon scattering amplitude (encoding a multitude of correlated scattering events) rather than the strength of a single paramagnon. 
This enhancement (quantified by $\mathrm{Re}\, \lambda_{\Mch} \!>\! 1$) is contributed by \textit{both} strong temporal correlations in the scattering (introduced by using the DMFT vertex as input) and the interplay of multiple nonlocal scattering channels (enabled by solving the parquet equations).
If one misses either of the two ingredients, the PG cannot be reproduced, as shown by the PA [cf.~\Fig{fig:Wu_PA_and_trilex}(b)~\cite{supplemental}] and $\ell$\DGA [cf.~\Fig{fig:Wu_self_energy}(c)] results, respectively.
Such a scattering enhancement mechanism has, to our best knowledge, not been seen before and sheds new light on the strong-coupling PG.

The fd approach is very general and can be applied with different nonperturbative input (e.g.\ HA or DMFT) and various levels of approximation in the field-theoretical solution (asymptotic or SBE decompositions, form-factor truncation, etc.). Thanks to this versatility, we anticipate many follow-up works, with cluster embedding~\cite{Slezak2009,Hafermann2008,Yang2011,Ayral2017,Iskakov2018,vanLoon2021,Iskakov2024,Kiese2024}, multi-orbital systems~\cite{Galler2017,Galler2019}, and real-frequency calculations~\cite{Kugler2021, Lee2021, Ge2024, Lihm2024, Ritz2024, Ritz2025} as promising directions.

\begin{acknowledgments}
The authors thank Yuan-Yao He for providing determinant quantum Monte Carlo results for \Fig{fig:highT}.
We thank Antoine Georges, Jason Kaye, Friedrich Krien, Thomas Sch\"afer, Alessandro Toschi, and Jan von Delft for helpful discussions.
D.K.\ thanks Olivier Parcollet and Michel Ferrero for ongoing collaborations on DiagMC.

J.-M.L.\ thanks Myungseok Nam for checking the consistency between the manuscript and the software implementation.
The NRG results were generated with the MuNRG package~\cite{Lee2016,Lee2017,Lee2021,Kugler2022} built on top of the QSpace tensor library~\cite{Weichselbaum2012a,Weichselbaum2020,Weichselbaum2024}.
This work was supported by the Fonds de la Recherche Scientifique - FNRS under Grants number T.0183.23 (PDR) and T.W011.23 (PDR-WEAVE).
Computational resources have been provided by the Consortium des Équipements de Calcul Intensif (CÉCI), funded by the FRS-FNRS under Grant No.~2.5020.11, and by the Tier-1 supercomputer of the Walloon Region (Lucia) with infrastructure funded by the Walloon Region under the Grant Agreement No.~1910247.
S.-S.B.L.~was supported by the National Research Foundation of Korea (NRF) grants funded by the Korean government (MSIT) (No.~RS-2023-00214464, No.~RS-2023-00258359, No.~RS-2023-NR119931, No.~RS-2024-00442710), the Global-LAMP Program of the NRF grant funded by the Ministry of Education (No.~RS-2023-00301976), the NRF grant funded by the Korean government (MEST) (No.~2019R1A6A1A10073437), and Samsung Electronics Co., Ltd.~(No.~IO220817-02066-01). F.B.K.\ acknowledges funding from the Ministerium f\"ur Kultur und Wissenschaft des Landes Nordrhein-Westfalen (NRW-R\"uckkehrprogramm).
The Flatiron Institute is a division of the Simons Foundation. 
\end{acknowledgments}

\textbf{Author contributions}
J.-M.L.\ and F.B.K.\ derived the analytic theory;
J.-M.L.\ and D.K.\ developed the code and performed the calculations;
all authors analyzed the results and wrote the paper.

\textbf{Competing interests}
The authors declare no competing interest.

\textbf{Data, Materials, and Software Availability}
The data and the codes used for this study are available at Ref.~\cite{fdDGAsolver}. All other data are included in the manuscript and/or SI Appendix.

\clearpage

\thispagestyle{empty}

\setcounter{equation}{0}
\setcounter{figure}{0}
\setcounter{table}{0}
\setcounter{page}{1}
\setcounter{section}{0}
\setcounter{secnumdepth}{2}

\renewcommand{\theequation}{S\arabic{equation}}
\renewcommand{\thefigure}{S\arabic{figure}}
\renewcommand{\thetable}{S\arabic{table}}
\renewcommand{\thepage}{S\arabic{page}}
\renewcommand{\thesection}{S-\Roman{section}}

\title{Supporting Information for ``The finite-difference parquet method: \\ Enhanced electron-paramagnon scattering opens a pseudogap''}

\date{March 7, 2026}
\maketitle

\section{Details of the formalism and implementation}

\subsection{Parametrization of the vertex function}
\begin{figure}
\centering

\includegraphics[width = 1\linewidth]{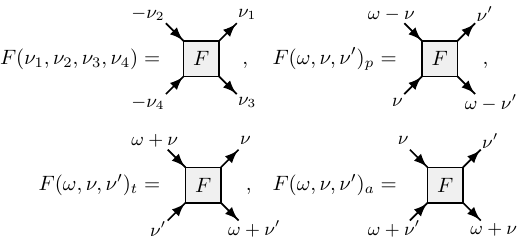}
\caption{Frequency conventions used in this work.}
\label{fig:conventions}
\end{figure}

Using spin-rotation invariance, we can parameterize the two-particle vertex as
\begin{align}
    F_{\sigma_1 \sigma_2 \sigma_3 \sigma_4} = F^{=} \delta_{\sigma_1 \sigma_2} \delta_{\sigma_3 \sigma_4} + F^{\times} \delta_{\sigma_1 \sigma_4} \delta_{\sigma_2 \sigma_3} \,,
\end{align}
where the choice of incoming and outgoing legs is shown in the first diagram of \Fig{fig:conventions}.
In the notation of Ref.~\cite{Rohringer2018}, we have
$F^{\uparrow \uparrow \uparrow \uparrow} = F^{\uparrow \uparrow} = F^= + F^\times$,
$F^{\uparrow \uparrow \downarrow \downarrow} = F^{\uparrow \downarrow} = F^=$,
and
$F^{\uparrow \downarrow \downarrow \uparrow} = F^{\overline{\uparrow\downarrow}} = F^{\times}$.
Vertices in the physical singlet (S), triplet (T), density (D), and magnetic (M) channels are given by
\begin{alignat}{2} \label{eq:physical_spin_F}
    & F_\Sch = F^= - F^{\times} \,,\ %
    &&F_\Tch = F^= + F^{\times} \,,\ %
    \nnnl
    & F_\Dch = 2F^= + F^{\times} \,,\ %
    &&F_\Mch = F^{\times} \,.
\end{alignat}
Accordingly, the bare vertex in each channel is given by
\begin{alignat}{3} \label{eq:bare_vertex}
    &F_0^=  = U \,,\ &&F_0^\times = -U \,,  &&F_\Sch = 2U \,,
    \nnnl
    &F_\Tch = 0 \,,\ &&F_\Mch     = -U \,,\ &&F_\Dch = U \,.
\end{alignat}

The vertex obeys a set of symmetry relations~\cite{Rohringer2018}:
\begin{subequations}
\begin{align}
    \label{eq:symm_cross_1}
    F^{=/\times}(1, 2, 3, 4)
    &= -F^{\times/=}(3, 2, 1, 4) \\
    \label{eq:symm_cross_2}
    &= -F^{\times/=}(1, 4, 3, 2) \\
    \label{eq:symm_1}
    &= F^{=/\times}(3, 4, 1, 2) \\
    \label{eq:symm_2}
    &= F^{=/\times}(4, 3, 2, 1) \\
    \label{eq:symm_3}
    &= F^{=/\times}(2, 1, 4, 3) \,.
\end{align}
\end{subequations}
Here, we use numeric arguments to denote frequencies (e.g., $1 = \nu_1$) or composite frequency-momentum variables (e.g., $1 = (\nu_1, \mb{k}_1) = k_1$).
Equations \eqref{eq:symm_cross_1} and \eqref{eq:symm_cross_2} indicate the crossing symmetries for the exchange of the two outgoing or incoming legs, respectively.
\EQ{eq:symm_1} comes from the combination of both,
\Eq{eq:symm_2} from complex conjugation and time reversal, and \Eq{eq:symm_3} from the combination of \Eqs{eq:symm_1} and \eqref{eq:symm_2}.
Thanks to these symmetries, it suffices to explicitly compute and store only one of the two spin components.
We therefore drop the $= / \times$ superscript in the following and simply write $F^{=} = F$, unless specified otherwise.

By energy conservation, $F$ depends on only three frequencies.
We parametrize it using two fermionic frequencies, $\nu$ and $\nu'$, and one bosonic frequency, $\omega$, and adopt a specific convention for each diagrammatic channel:
\begin{subequations} \label{eq:freq_convention}
\begin{align}
    \label{eq:freq_convention_p}
    F(\omega, \nu, \nu')_{p}
    &= F(\nu',\ \nu - \omega, \ \omega - \nu',  \ -\nu)
    \,, \\
    \label{eq:freq_convention_t}
    F(\omega, \nu, \nu')_{t} &= F(\nu,\ -\nu - \omega,\ \nu' + \omega,\ -\nu')
    \,, \\
    \label{eq:freq_convention_a}
    F(\omega, \nu, \nu')_{a} &= F(\nu',\ -\nu,\ \nu + \omega,\ -\nu' - \omega)
    \,.
\end{align}
\end{subequations}
These conventions are illustrated in \Fig{fig:conventions}.
For the $r$-reducible and -irreducible vertices $\Phi_r$ and $I_r$, we always use the $r$-channel convention.
With explicit frequency arguments, \Eq{eq:F_parquet} (parametrized in the $t$-channel) reads
\begin{align} \label{eq:F_fd}
    &F(\omega, \nu, \nu')_{t}
    = \Lambda(\omega, \nu, \nu')_{t}
    + \Phi_t(\omega, \nu, \nu')
    \nnnl
    &+ \Phi_a(\nu'-\nu, \omega + \nu, \nu)
    + \Phi_p(\nu+\nu'+\omega, \nu', \nu) \,.
\end{align}
Our convention for the frequency notation follows Ref.~\cite{Kiese2024a}.
We choose the same parametrization for momentum variables, using $\mb{k}$ and $\mb{k'}$ for fermionic and $\mb{q}$ for bosonic momenta.
We denote composite frequency-momentum variables as $k = (\nu, \mb{k})$ and $q = (\omega, \mb{q})$.

The frequency structure of the reducible vertices $\Phi$ can be accurately captured using the asymptotic decomposition pioneered by Li \textit{et al.}~\cite{Li2016} and Wentzell \textit{et al.}~\cite{Wentzell2020},
\begin{align} \label{eq:vertex_asymptotics}
    \Phi_r(\omega, \nu, \nu')
    &= K_{1r}(\omega) + K_{2r}(\omega, \nu) + K_{\bar{2}r}(\omega, \nu') \nnnl
    &\ + K_{3r}(\omega, \nu, \nu') \,,
\end{align}
where we use a subscript $\bar{2}$ instead of the standard choice $2'$ for compactness.
For local interactions, $K_{1r}$ and $K_{2r} / K_{\bar{2}r}$ are closely related to physical two- and three-point response functions~\cite{Wentzell2020, Tagliavini2019} and capture the high-frequency behavior of $\Phi_r$.

\subsection{Bethe--Salpeter equations}

The channel-specific parametrizations make the BSE diagonal in the corresponding bosonic frequency.
Defining the bubble functions
\begin{subequations} \label{eq:BSE_channels_explicit}
\begin{align}
    \Pi_p(\omega, \nu) &= G(\nu) G(\omega-\nu) \,, \\
    \Pi_a(\omega, \nu) = \Pi_t(\omega, \nu) &= G(\nu) G(\omega+\nu) \,,
\end{align}
\end{subequations}
the BSEs read
\begin{subequations} \label{eq:BSE_pta}
\begin{align} \label{eq:BSE_p}
    \Phi^=_p(\omega, \nu, \nu')
    &= \frac{1}{\beta} \sum_{\tilde{\nu}}
    I^{=}_p(\omega, \nu, \tilde{\nu})
    \Pi_p(\omega, \tilde{\nu})
    F^{\times}(\omega, \tilde{\nu}, \nu')_p
    \,,
\end{align}
\begin{align} \label{eq:BSE_t}
    \Phi^=_t(\omega, \nu, \nu')
    &= \frac{2}{\beta} \sum_{\tilde{\nu}}
    I^{=}_t(\omega, \nu, \tilde{\nu}) \,
    \Pi_t(\omega, \tilde{\nu}) \,
    F^{=}(\omega, \tilde{\nu}, \nu')_t
    \nnnl
    &+ \frac{1}{\beta} \sum_{\tilde{\nu}}
    I^{=}_t(\omega, \nu, \tilde{\nu}) \,
    \Pi_t(\omega, \tilde{\nu}) \,
    F^{\times}(\omega, \tilde{\nu}, \nu')_t
    \nnnl
    &+ \frac{1}{\beta} \sum_{\tilde{\nu}}
    I^{\times}_t(\omega, \nu, \tilde{\nu}) \,
    \Pi_t(\omega, \tilde{\nu}) \,
    F^{=}(\omega, \tilde{\nu}, \nu')_t
    \,,
\end{align}
\begin{align} \label{eq:BSE_a}
    \Phi^=_a(\omega, \nu, \nu')
    &= -\frac{1}{\beta} \sum_{\tilde{\nu}}
    I^{=}_a(\omega, \nu, \tilde{\nu}) \,
    \Pi_a(\omega, \tilde{\nu}) \,
    F^{=}(\omega, \tilde{\nu}, \nu')_a
    \,.
\end{align}
\end{subequations}
We explicitly write the spin components as the BSEs in the $p$ and $t$ channels mix the two spin components.

To diagonalize the $t$-channel BSE in the spin index, one uses the physical M and D channels, with
\begin{equation} \label{eq:physical_spin_Phi}
    \Phi_{\Mch} = \Phi_t^{\times} \,, \quad
    \Phi_{\Dch} = 2\Phi_t^{=} + \Phi_t^{\times} \,,
\end{equation}
and similarly the irreducible vertices $I_\Mch$, $I_\Dch$.
By crossing symmetry, the $a$-channel reducible vertex is obtained as
\begin{equation} \label{eq:physical_spin_Phi_cross}
    \Phi_a^=           = -\Phi_t^{\times} = -\Phi_\Mch \,,\quad
    \Phi_a^{\times} = -\Phi_t^=           = \tfrac{1}{2} \Phi_\Mch - \tfrac{1}{2} \Phi_\Dch  \,.
\end{equation}
In \Eqs{eq:physical_spin_Phi} and \eqref{eq:physical_spin_Phi_cross}, we omitted the argument $(\omega, \nu, \nu')$ for brevity.
Then, the full vertex in, e.g., the M channel reads
\begin{align} \label{eq:F_fd_spin_basis}
    F_\Mch(\omega, \nu, \nu')_{t}
    & = \Lambda_\Mch(\omega, \nu, \nu')_{t} + \Phi_\Mch(\omega, \nu, \nu')
    \nnnl
    &+ \tfrac{1}{2} \Phi_{\overbar{\Mch}}(\omega, \nu, \nu') - \tfrac{1}{2} \Phi_{\overbar{\Dch}}(\omega, \nu, \nu') 
    \nnnl
    &+ \Phi_p^{\times}(\nu+\nu'+\omega, \nu', \nu)
    \,,
\end{align}
where we defined the transverse magnetic ($\overbar{\rm M}$) and transverse density ($\overbar{\rm D}$) vertices as
\begin{equation} \label{eq:transverse_MD_def}
    \Phi_{\overbar{\Mch}\scriptscriptstyle{/}\overbar{\Dch}}(\omega, \nu, \nu') =
    \Phi_{\MDch}(\nu'-\nu,\ \omega+\nu,\ \nu) \,,
\end{equation}
to account for the different frequency parametrizations.
Since we define the M and D channels with the $t$-channel parametrization, the transverse M and D channels follow the $a$-channel parametrization.
As desired, the BSE is diagonal in the M and D channels:
\begin{multline} \label{eq:BSE_MD}
    \Phi_{\MDch}(\omega, \nu, \nu')
    \\
    = \frac{1}{\beta} \sum_{\tilde{\nu}}
    I_{\MDch}(\omega, \nu, \tilde{\nu})
    \Pi_{t}(\omega, \tilde{\nu})
    F_{\MDch}(\omega, \tilde{\nu}, \nu')_t \,.
\end{multline}
For the $p$ channel, we can keep using $\Phi_p^=$ by exploiting crossing symmetry to rewrite \Eq{eq:BSE_p} as
\begin{equation}
    \Phi^{=}_{p}(\omega, \nu, \nu')
    = -\frac{1}{\beta} \sum_{\tilde{\nu}}
    I^{=}_p(\omega, \nu, \tilde{\nu})
    \Pi_p(\omega, \tilde{\nu})
    F^{=}(\omega, \omega - \tilde{\nu}, \nu')_p \,.
\end{equation}

For lattice systems, the frequency summation in the BSE is augmented by a momentum summation $\frac{1}{N_\mb{k}} \sum_\mb{k}$, where $N_\mb{k}$ is the number of discretized momentum points.
The nonlocal $a$-channel BSE, e.g., reads
\begin{equation}
    \Phi^{=}_{a}(q, k, k')
    = -\sum_{\tilde{k}}
    I^{=}_a(q, k, \tilde{k})
    \Pi_a(q, \tilde{k})
    F^{=}(q, \tilde{k}, k')_a \,,
\end{equation}
where $\sum_{\tilde{k}} = \frac{1}{\beta N_\mb{k}} \sum_{\tilde{\nu}, \tilde{\mb{k}}}$.

In this work, we use the $s$-wave approximation for the reducible vertices~\cite{Lichtenstein2017, Eckhardt2018}, keeping their dependence on the bosonic momentum $\mb{q}$ but discarding their dependence on $\mb{k}$ and $\mb{k'}$~\cite{Lichtenstein2017, Eckhardt2018, Eckhardt2020}.
Within this approximation, the nonlocal $a$-channel BSE, e.g., becomes
\begin{multline}
    \Phi^{=,\,\swave}_{a}(q, \nu, \nu')
    = -\frac{1}{\beta}\sum_{\tilde{\nu}}
    I^{=,\,\swave}_a(q, \nu, \tilde{\nu})
    \\
    \times \Pi^\swave_a(q, \tilde{\nu})
    F^{=,\,\swave}_a(q, \tilde{\nu}, \nu') \,,
\end{multline}
where the $s$-wave vertex and bubble functions are 
\begin{subequations} \label{eq:swave_vertex_bubble}
\begin{align}
    F^{=,\,\swave}_a(q, \nu, \nu')
    &= \frac{1}{N_\mb{k}^2} \sum_{\mb{k}, \mb{k'}} F^{=}_a(q, k, k') \,,
    \\
    \Pi^{\swave}_a(q, \nu)
    &= \frac{1}{N_\mb{k}} \sum_{\mb{k}} \Pi_a(q, k) \,.
\end{align}
\end{subequations}

Finally, the susceptibility $\chi_\MDch(q)$, bosonic propagator $w_\MDch(q)$, and Hedin vertex $\lambda_\MDch(q, \nu)$ can be expressed through the asymptotic vertex functions as~\cite{Bonetti2022}
\begin{subequations} \label{eq:lDGA_chi_and_hedin}
\begin{align}
    \chi_{\MDch}(q) &= \frac{K_{1,{\MDch}}(q)}{U^2} \,,
    \\
    w_{\MDch}(q) &= K_{1,\MDch}(q) \mp U \,,
    \\
    \lambda^\swave_{\MDch}(q, \nu) &= 1 + \frac{K^\swave_{2,{\MDch}}(q, \nu)}{K_{1,{\MDch}}(q) \mp U} \,.
\end{align}
\end{subequations}
Due to the $s$-wave approximation, the Hedin vertex depends only on the bosonic but not the fermionic momentum.
The Hedin vertex then satisfies~\cite{Krien2019b, Krien2021}
\begin{multline} \label{eq:hedin_bse_swave}
    \lambda^\swave_\MDch(q, \nu)
    = 1
    + \frac{1}{\beta} \sum_{\nu'} \Bigl[ F^{\swave}_{\MDch}(q, \nu, \nu')_t
    \\
    - S^\swave_\MDch(q, \nu, \nu') \Bigr]
    \Pi^{\swave}_t(q, \nu') \,,
\end{multline}
with the $U$-reducible vertex
\begin{equation}
    S^\swave_\MDch(q, \nu, \nu')
    = \lambda^\swave_\MDch(q, \nu) w_\MDch(q) \lambda^\swave_\MDch(q, \nu') \,.
\end{equation}

\subsection{Fd equations in the asymptotic and single-boson exchange parametrizations}

In the main text, we derived the fd-BSE \eqref{eq:fdBSE} for the reducible vertices $\tilde{\Phi}_r$.
Here, we derive the corresponding equations for $\tilde{\Phi}_r$ in the asymptotic and single-boson exchange parametrizations.

We begin with the asymptotic parametrization of \Eq{eq:vertex_asymptotics}~\cite{Wentzell2020}.
We define~\cite{Gievers2022}
\begin{subequations}
\begin{align}
F_{012r} & = F_0 + K_{1r} + K_{2r}
, \\
F_{01\bar{2}r} & = F_0 + K_{1r} + K_{\bar{2}r}
,
\end{align}
\end{subequations}
the parts of $F$ depending on only $\omega$ and $\nu$, or $\omega$ and $\nu'$, respectively.
We then write the BSE~\eqref{eq:BSE} as
\begin{subequations}
\begin{alignat}{2}
K_{1r} & = F_0 && \Pi_r F_{012r}
, \\
\label{eq:BSE_K2}
K_{2r} & = (I_r-F_0) && \Pi_r F_{012r}
, \\
K_{\bar{2}r} & = F_0 && \Pi_r (F-F_{012r})
, \\
\label{eq:BSE_K3}
K_{3r} & = (I_r-F_0) && \Pi_r (F-F_{012r})
.
\end{alignat}
\end{subequations}
It is easy to see that the sum of these four equations reproduces \Eq{eq:BSE}
and that the restricted frequency dependencies of $K_{1r}$, $K_{2r}$, and $K_{\bar{2}r}$ are fulfilled.

We further split off the factorizable three-dimensional frequency dependence of $K_{3r}$. To this end, we first write
(see, e.g., Eq.~(80) in Ref.~\cite{Gievers2022})
\begin{align}
K_{3r} & = M_r + K_{2r} W_r^{-1} K_{\bar{2}r}
, \quad
W_r = F_0 + K_{1r}
\label{eq:K3_M_K2WK2}
\end{align}
where $W_r$ is a boson propagator and $M_r$ the multiboson exchange part~\cite{Krien2019b}.
The single-boson exchange part is
\begin{subequations}
\begin{align}
S_r & = F_0 + K_{1r} + K_{2r} + K_{\bar{2}r} + K_{2r} W_r^{-1} K_{\bar{2}r}
\\
& =
(1 + K_{\bar{2}r} W_r^{-1}) W_r (1 + W_r^{-1} K_{2r})
,
\end{align}
\end{subequations}
where we write $S_r$ instead of the usual $\nabla_r$~\cite{Krien2019b} to allow for upper- and lower-case notation.
Using the relations
\begin{subequations}
\label{eq:F012_both}
\begin{align}
\label{eq:F012}
F_{012r} W_r^{-1} K_{\bar{2}r}
& =
K_{\bar{2}r} + K_{2r} W_r^{-1} K_{\bar{2}r}
= S_r - F_{012r}
,
\\
\label{eq:F012p}
K_{2r} W_r^{-1} F_{01\bar{2}r} 
& =
K_{2r} + K_{2r} W_r^{-1} K_{\bar{2}r}
= S_r - F_{01\bar{2}r}
,
\end{align}
\end{subequations}
we can write the BSE~\eqref{eq:BSE_K3} for $K_{3r}$ as
\begin{subequations}
\begin{align}
K_{3r} & = (I_r-F_0) \Pi_r (F-S_r
+ F_{012r} W_r^{-1} K_{\bar{2}r})
\\
& = 
(I_r-F_0) \Pi_r (F-S_r)
+ K_{2r} W_r^{-1} K_{\bar{2}r}
,
\end{align}
\end{subequations}
where we used Eq.~\eqref{eq:BSE_K2} in the last step.
One obtains the desired BSE for the multiboson exchange part~\cite{Krien2021},
\begin{align}
M_r = (I_r - F_0) \Pi_r (F - S_r)
.
\end{align}

We now move to the finite-difference BSEs~\eqref{eq:fdBSE}.
The asymptotic classes obey
\begin{subequations}
\label{eq:fd_BSE_K}
\begin{alignat}{2}
\label{eq:fd_BSE_K1}
\tilde{K}_{1r} & = f_{01\bar{2}r} && (\tilde{\Pi}_r + \pi_r \tilde{I}_r \Pi_r) F_{012r}  
, \\
\tilde{K}_{2r} & = (f-f_{01\bar{2}}) && (\tilde{\Pi}_r + \pi_r \tilde{I}_r \Pi_r) F_{012r}
\nonumber \\
& \
+ \tilde{I}_r \Pi_r F_{012r} 
, && 
\label{eq:fd_BSE_K2}
\\
\tilde{K}_{\bar{2}r} & = f_{01\bar{2}r} && (\tilde{\Pi}_r + \pi_r \tilde{I}_r \Pi_r) (F-F_{012r})
\nonumber \\
& &&
+ f_{012r} \pi_r \tilde{I}_r
, 
\label{eq:fd_BSE_K2p}
\\
\tilde{K}_{3r} & = (f-f_{01\bar{2}r}) && (\tilde{\Pi}_r + \pi_r \tilde{I}_r \Pi_r) (F-F_{012r})
\nonumber \\
& \
+ \tilde{I}_r \Pi_r (F-F_{012r})
&& + (f-f_{012r}) \pi_r \tilde{I}_r
.
\label{eq:fd_BSE_K3}
\end{alignat}
\end{subequations}
Once more, it is easy to see that the sum of these equations reproduces Eq.~\eqref{eq:fdBSE}.

We again split off the factorizable three-dimensional frequency dependence of $\tilde{K}_{3r}$. 
From Eq.~\eqref{eq:K3_M_K2WK2}, we have
\begin{align} \label{eq:fd_BSE_K3_split}
\tilde{K}_{3r} 
& = 
\tilde{M}_r 
+ 
\tilde{K}_{2r} W_r^{-1} K_{\bar{2}r}
\nonumber \\
& \ 
- 
k_{2r} w_r^{-1} \tilde{W}_{r} W_r^{-1} K_{\bar{2}r}
+ 
k_{2r} w_r^{-1} \tilde{K}_{\bar{2}r}
,
\end{align}
where we used 
$W_r^{-1} - w_r^{-1} = - w_r^{-1} (W_r - w_r) W_r^{-1}$,
and the fd multiboson exchange part is
\begin{align}
\tilde{M}_r
& =
(f-s_r) \hat{\Pi}_r (F-S_r)
\nonumber \\
& \
+ \tilde{I}_r \Pi_r (F-S_r)
+ (f-s_r) \pi_r \tilde{I}_r
,
\label{eq:fd_BSE_M}
\end{align}
where $\hat{\Pi}_r = \tilde{\Pi}_r + \pi_r \tilde{I}_r \Pi_r$ is a shorthand.
To verify \Eq{eq:fd_BSE_M}, we rephrase the first (second) line of \Eq{eq:fd_BSE_K3} using the first (second) lines of Eqs.~\eqref{eq:fd_BSE_K2}, \eqref{eq:fd_BSE_K2p}, and \eqref{eq:fd_BSE_M}, with line number specified by superscript $(1)$ [$(2)$].
The first line of \Eq{eq:fd_BSE_K3} is recast as
\begin{align}
\tilde{K}_{3r}^{(1)} 
\overset{\eqref{eq:F012p}}&{=} (f-s_r+k_{2r}w_r^{-1}f_{01\bar{2}r}) \hat{\Pi}_r (F-F_{012r})
\nnnl
\overset{\eqref{eq:fd_BSE_K2p}}&{=}
(f-s_r) \hat{\Pi}_r
(F-F_{012r})
+
k_{2r} w_r^{-1} \tilde{K}_{\bar{2}r}^{(1)}
\nnnl
\overset{\eqref{eq:F012}}&{=} 
(f-s_r) \hat{\Pi}_r 
(F-S_r+F_{012r}W_r^{-1}K_{\bar{2}r})
\nnnl
& \qquad + k_{2r} w_r^{-1} \tilde{K}_{\bar{2}r}^{(1)}
\nnnl
\overset{\eqref{eq:fd_BSE_M}}&{=} 
(f-s_r) \hat{\Pi}_r
F_{012r}W_r^{-1}K_{\bar{2}r}
\nnnl
& \qquad + k_{2r} w_r^{-1} \tilde{K}_{\bar{2}r}^{(1)}
+ 
\tilde{M}_r^{(1)}
\nnnl
\overset{\eqref{eq:F012}}&{=} 
(f-f_{01\bar{2}r}-k_{2r} w_r^{-1} f_{01\bar{2}r}) \hat{\Pi}_r
F_{012r}W_r^{-1}K_{\bar{2}r}
\nnnl
& \qquad +
k_{2r} w_r^{-1} \tilde{K}_{\bar{2}r}^{(1)}
+ 
\tilde{M}_r^{(1)}
\nnnl
\overset{\eqref{eq:fd_BSE_K1}, \eqref{eq:fd_BSE_K2}}&{=} 
-k_{2r} w_r^{-1} 
\tilde{W}_r
W_r^{-1}K_{\bar{2}r}
+
\tilde{K}_{2r}^{(1)}
W_r^{-1} K_{\bar{2}r}
\nnnl
& \qquad
+
k_{2r} w_r^{-1} \tilde{K}_{\bar{2}r}^{(1)}
+ 
\tilde{M}_r^{(1)}
,
\label{eq:fd_BSE_K3_line1}
\end{align}
where we also used $F_0 = f_0$ so that $\tilde{W}_r = \tilde{K}_{1r}$ for the last equality.
For the second line of \Eq{eq:fd_BSE_K3}, we have
\begin{align} \label{eq:fd_BSE_K3_line2}
& \tilde{K}_{3r}^{(2)} \overset{\eqref{eq:F012_both}}{=} 
\tilde{I}_r \Pi_r (F-S_r-F_{012r}W_r^{-1}K_{\bar{2}r})
\nnnl
& \qquad \qquad
+ (f-s_r+k_{2r}w_r^{-1}f_{01\bar{2}r}) \pi_r \tilde{I}_r
\nnnl
& \overset{\eqref{eq:fd_BSE_K2},\eqref{eq:fd_BSE_K2p},\eqref{eq:fd_BSE_M}}{=}
\tilde{M}_r^{(2)}
+
\tilde{K}_{2r}^{(2)} W_r^{-1} K_{\bar{2}r} + k_{2r} w_r^{-1} \tilde{K}_{\bar{2}r}^{(2)}
.
\end{align}
Adding \Eqs{eq:fd_BSE_K3_line1} and \eqref{eq:fd_BSE_K3_line2} and comparing it with the right side of \Eq{eq:fd_BSE_K3_split} yields \Eq{eq:fd_BSE_M}.

\subsection{Schwinger--Dyson equation}

The SDE, which relates the full two-particle vertex $F$ to the one-particle self-energy $\Sigma$, reads
\begin{equation} \label{eq:SDE}
    \Sigma(k) - \Sigma_\Hart
    = \mathrm{SDE}[F]\,.
\end{equation}
Here, $\Sigma_\Hart = U \langle n \rangle$ is the Hartree self-energy, and we defined the SDE functional for a vertex $A$ and a given fermionic propagator $G$ as 
\begin{equation}
    \mathrm{SDE}[A]
    = -U \sum_{k' q} A(q, k, k')_t G(k+q) G(k') G(k'+q)\,,
\end{equation}
using the $t$ channel parameterization [\Eq{eq:freq_convention_t}].
Importantly, the SDE functional is linear in the vertex argument.
As before, $F$ implicitly refers to $F^=$.

In the $s$-wave approximation, as used here, the $r$-reducible vertex loses its momentum dependence when converted to different channels and is thus best represented in its natural channel $r$.
Following Ref.~\cite{Hille2020}, we rewrite the SDE accordingly, using the $r$-channel representation for each $\Phi_r$. The $t$-channel contribution reads
\begin{subequations} \label{eq:SDE_decomp}
\begin{align} \label{eq:SDE_decomp_t}
    &\Sigma_t(k)
    = \mathrm{SDE} \bigl[ \Phi_t \bigr]
    \nnnl
    &= -U \sum_{k' q} \Phi_t(q, k, k') G(k+q) G(k') G(k'+q)
    \nnnl
    &= -U \sum_q G(k+q) \Bigl( \sum_{k'} \Phi_t(q, k, k') \Pi_t(q, k') \Bigr)
    \nnnl
    &= -\sum_q G(k+q) \Bigl( \Phi_t \Pi_t U \Bigr)(q, k) \,.
\end{align}
Similarly, the $a$ and $p$ channel contributions are given by
\begin{align}
    \label{eq:SDE_decomp_a}
    \Sigma_a(k)
    &= \mathrm{SDE} \bigl[ \Phi_a \bigr] \nnnl
    &= -\sum_q G(k+q) \Bigl( \Phi_a \Pi_a U \Bigr)(q, k)\,,
    \\
    \label{eq:SDE_decomp_p}
    \Sigma_p(k)
    &= \mathrm{SDE} \bigl[ \Phi_p \bigr] \nnnl
    &= -\sum_q G(q-k) \Bigl( \Phi_p \Pi_p U \Bigr)(q, k)\,.
\end{align}
\end{subequations}
Adding the fully irreducible vertex contribution gives
\begin{equation} \label{eq:sde_channels}
    \Sigma(k) - \Sigma_\Hart
    = \mathrm{SDE} [ \Lambda ]
    + \Sigma_a(k) + \Sigma_p(k) + \Sigma_t(k) \,.
\end{equation}

For fd-\DGA, we further apply the fd decomposition of the vertex, $F = f + \sum_{r=a,p,t} \tilde{\Phi}_r$.
We thus use
\begin{equation} \label{eq:sde_channels_fd}
    \Sigma(k) - \Sigma_\Hart
    = \mathrm{SDE} [ f ]
    + \tilde{\Sigma}_a(k) + \tilde{\Sigma}_p(k) + \tilde{\Sigma}_t(k) \,,
\end{equation}
where $\tilde{\Sigma}_r(k)$ are defined as in Eqs.~\eqref{eq:SDE_decomp} but using the fd vertices,
$\tilde\Sigma_r(k) = \mathrm{SDE} \bigl[ \tilde\Phi_r \bigr]$.
Note that 
$\tilde{\Sigma}_a(k) + \tilde{\Sigma}_p(k) + \tilde{\Sigma}_t(k)
=
\tilde{\Sigma}_\Mch(k) + \tilde{\Sigma}_\Dch(k) + \tilde{\Sigma}_\Sch(k)$
(as used in \Fig{fig:Wu_diagnostics}) with
\begin{equation}
    \tilde{\Sigma}_\Mch = \frac{3}{2} \tilde{\Sigma}_a \,,\ %
    \tilde{\Sigma}_\Dch = \tilde{\Sigma}_t - \frac{1}{2} \tilde{\Sigma}_a \,,\ %
    \tilde{\Sigma}_\Sch = \tilde{\Sigma}_p \,.
\end{equation}
The triplet channel does not contribute to the self-energy due to crossing symmetry.

\subsection{Postprocessing correction for $K_1$ and $\Sigma$}
Due to the $s$-wave approximation, the coefficient of the $1/\nu$ tail of the self-energy is not correctly captured in our p\DGA calculations for the HM.
In the main text, we corrected this problem through the ``local correction'' \Eq{eq:local_corr}, setting the local component of the self-energy to the DMFT result.
In this section, we discuss an alternative strategy to correct the self-energy tail.

The high-frequency asymptote of the self-energy is
\begin{equation} \label{eq:tail_selfen}
    \Sigma(k)
    = \Sigma_\Hart + \frac{U^2 n_\sigma (1-n_\sigma)}{i\nu} + O\left( \frac{1}{\nu^2} \right) \,,
\end{equation}
with $n_\sigma$ the density per spin.
It traces back~\cite{Chalupa2022} to the sum rule of the equal-spin susceptibility
\begin{equation} \label{eq:tail_chi_sumrule}
    \sum_q \chi_{\sigma\sigma}(q)
    = n_\sigma (1 - n_\sigma) \,,
\end{equation}
which is fulfilled if $\chi_{\sigma\sigma}$ is deduced from a crossing-symmetric vertex~\cite{Chalupa2022}.
The equal-spin susceptibility can be expressed through $K_1$ in the physical channels via
\begin{equation}
    \label{eq:chi_def1}
    \chi_{\MDch}(q) = \frac{K_{1\MDch}(q)}{U^2}
    = \chi_{\sigma\sigma}(q) \mp \chi_{\sigma\bar{\sigma}}(q)
    \,.
\end{equation}

\begin{figure}[tb]
\centering
\includegraphics[width = 0.99\linewidth]{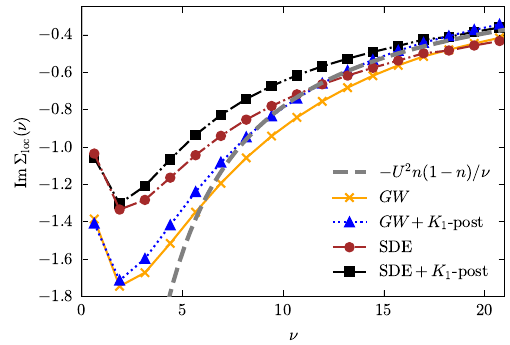}
\caption{Correction of the local self-energy $\Sigma_{\rm loc}(\nu) = \frac{1}{N_\mb{k}} \sum_\mb{k} \Sigma(\nu, \mb{k})$ of the HM (parameters same as \Fig{fig:Wu_self_energy}).
While the $GW$ self-energy $\Sigma^{\GW} [ K_1 ]$ obtained directly from the $s$-wave-approximated p\DGA [yellow crosses; cf.~\Eq{eq:tail_GW}] deviates from the exact asymptotics $\im \Sigma \sim -U^2 n_\sigma(1-n_\sigma) / \nu$ (gray dashed curve), that from the postprocessed $K_1$ vertex $\Sigma^{\GW} [ K_1^\mr{post} ]$ [blue triangles; cf.~\Eq{eq:tail_K1_pp_full}] has the correct asymptotics.
By adding $\Sigma^{\GW} [ K_1^\mr{post} ] - \Sigma^{\GW} [ K_1 ]$ to the uncorrected SDE self-energy (red circles),
we obtain a corrected self-energy with the correct asymptotic behavior [\Eq{eq:tail_sigma_corr}] (black squares).
}
\label{fig:tail_correction}
\end{figure}

While the exact p\DGA is crossing symmetric, the $s$-wave approximation breaks this symmetry when applying the BSE~\cite{Eckhardt2020}, thus violating the $\chi_{\sigma\sigma}$ sum rule and $\Sigma$ asymptote.
To understand the violation of the self-energy asymptote, we inspect the large-$\nu$ limit of the SDE \eqref{eq:sde_channels}.
Using $G(k) = O(1/\nu)$, $\Lambda = U + O(1/\nu)$, and
\begin{subequations}
    \begin{align}
        \lim_{\nu \to \infty} \Bigl( (\Phi_p+U) \Pi_p U \Bigr)(q, k)
        &= -K_{1p}(q) \,,
        \\
        \lim_{\nu \to \infty} \Bigl( (\Phi_a+U) \Pi_a U \Bigr)(q, k)
        &= -K_{1a}(q) \,,
        \\
        \lim_{\nu \to \infty} \Bigl( (\Phi_t+U) \Pi_t U \Bigr)(q, k)
        &= K_{1t}(q) -K_{1a}(q) \,,
    \end{align}
\end{subequations}
we find that the large-$\nu$ limit of the self-energy is
\begin{equation}
    \Sigma(k) - \Sigma_\Hart
    = {\rm SDE}\bigl[ F \bigr](k)
    = \Sigma^\GW(k) + O\Bigl(\frac{1}{\nu^2}\Bigr) \,,
    \label{eq:tail_sigma_nu-2}
\end{equation}
using the $GW$ and second-order expressions
\begin{align} 
    \Sigma^\GW(k)
    & = \sum_q \Bigl[
    G(k+q) \bigl( 2K_{1a}(q) - K_{1t}(q) \bigr)
    \nonumber \\
    & \quad 
    + G(q-k) K_{1p}(q)
    \Bigr] - 2\Sigma^{U^2}(k) \,,
    \label{eq:tail_GW}
    \\
    \Sigma^{U^2}(k) & = -U^2 \sum_{k', q} G(k') G(q-k') G(q-k)
    \label{eq:tail_PT2}
    .
\end{align}
Since $\Sigma^{U^2}(k)$ does not depend on the vertex, the erroneous tail must stem from inaccurate $K_1$ in \Eq{eq:tail_GW}.

To mitigate this problem, we recompute the $K_1$ vertices using the postprocessing (post) formula
\begin{equation} \label{eq:tail_K1_pp}
    K_{1r}^{\rm post} = U \Pi_r U + U \Pi_r F \Pi_r U \,,
\end{equation}
which is also used in functional renormalization group (fRG) calculations~\cite{Tagliavini2019, Hille2020, Chalupa2022}.
The postprocessed and directly calculated $K_1$ vertices are identical if the \DGA equations are solved exactly.
However, in practice, the two deviate due to the numerical approximations involved, in particular, the $s$-wave approximation in $\Phi_r$.
The benefit of the postprocessed $K_1$ is that they exactly satisfy the equal-spin susceptibility sum rule \eqref{eq:tail_chi_sumrule} given a crossing-symmetric $F$~\cite{Hille2020, Chalupa2022}, which is the case for $s$-wave approximated p\DGA.
Writing momentum dependences explicitly, the postprocessed $K_{1a}$ reads
\begin{align} \label{eq:tail_K1_pp_full}
    K_{1a}^{\rm post}(q)
    &= U^2 \sum_k \Pi_a(q, k) \Bigl[ 1 + \sum_{k'} F(q, k, k')_a \Pi_a(q, k') \Bigr]  \,.
\end{align}
Similar equations can be found for the $t$ and $p$ channels using the BSE \eqref{eq:BSE_channels_explicit}.
The $GW$ self-energy \eqref{eq:tail_GW} evaluated with the difference between the postprocessed and direct $K_1$ vertex can be added to correct the self-energy,
\begin{equation} \label{eq:tail_sigma_corr}
    \Sigma^{\rm post}
    = \Sigma
    + \Sigma^\GW\bigl[ K_1^{\rm post} \bigr]
    - \Sigma^\GW\bigl[ K_1 \bigr] \,.
\end{equation}
Note that the contributions to $\Sigma$ from $K_2$, $K_3$, and the two-particle irreducible vertices decay faster than $1/\nu$ [cf.~Eqs.~\eqref{eq:tail_sigma_nu-2}--\eqref{eq:tail_PT2}], so the $1/\nu$ tail of $\Sigma^\mr{post}$ is governed by the post-processing.

\begin{figure}[tb]
    \centering
    \includegraphics[width = 0.99\linewidth]{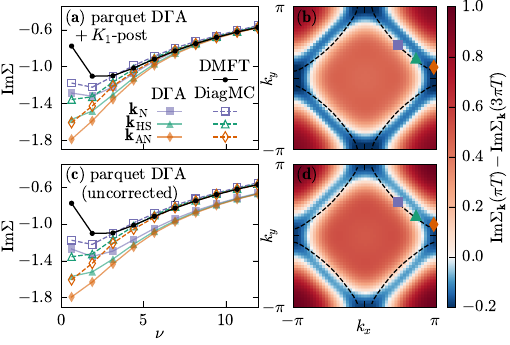}
    \caption{
    Same as \Fig{fig:Wu_self_energy}, but comparing p\DGA (a), (b) with the $K_1$-postprocessing correction [\Eq{eq:tail_sigma_corr}], and (c), (d) without any correction.
    }
    \label{fig:Wu_tail_compare}
\end{figure}

Figures~\ref{fig:tail_correction} and \ref{fig:Wu_tail_compare} illustrate that this $K_1$-postprocessing correction scheme gives a rigid, momentum-independent shift of the high-frequency tail and weakly affects the low-frequency self-energy.
The hierarchy of strongly insulating antinode (AN), weakly insulating hot spot (HS), and metallic node (N) remains unaffected.

\subsection{Ladder \DGA}

For completeness, we here briefly summarize the fd formulation of \lDGA~\cite{Galler2017} and the explicit equations in the asymptotic decomposition of the vertex.
Within \lDGA, the local vertex (typically computed from DMFT) is first decomposed into irreducible and reducible parts in the magnetic (M) and density (D) channels:
\begin{equation}
    f_{\MDch}(\omega, \nu, \tilde{\nu})_t = i_{\MDch}(\omega, \nu, \tilde{\nu}) + \phi_{\MDch}(\omega, \nu, \tilde{\nu}) \,.
\end{equation}
Then, one assumes the M/D irreducible vertex to be identical between the impurity and the lattice.
This yields the BSE for the nonlocal reducible vertex [see \Eq{eq:BSE_MD}]:
\begin{multline} \label{eq:lDGA_BSE}
    \Phi_{\MDch}(q, \nu, \tilde{\nu})
    = \frac{1}{\beta} \sum_{\tilde{\nu}} i_\MDch(\omega, \tilde{\nu}, \nu') \, \Pi_t(q, \tilde{\nu})
    \\
    \times \bigl[ i_{\MDch}(\omega, \tilde{\nu}, \nu') + \Phi_{\MDch}(q, \tilde{\nu}, \nu') \bigr] \,.
\end{multline}
Taking steps similar to Eqs.~\eqref{eq:fdBSE_step1}--\eqref{eq:fdBSE}, we find
the fd BSE for the fd vertices $\tilde{\Phi}_{\MDch} = \Phi_{\MDch} - \phi_{\MDch}$, which reads~\cite{Galler2017}
\begin{multline} \label{eq:lDGA_fdBSE}
    \tilde{\Phi}_{\MDch}(q, \nu, \nu')
    =
    \frac{1}{\beta} \sum_{\tilde{\nu}} f_{\MDch}(\omega, \nu, \tilde{\nu})_t \,
    \tilde{\Pi}_t(q, \tilde{\nu})
    \\
    \times \bigl[ f(\omega, \tilde{\nu}, \nu')_t + \tilde{\Phi}_{\MDch}(q, \tilde{\nu}, \nu') \bigr] \,,
\end{multline}
with the fd bubble function
\begin{equation}
    \tilde{\Pi}_t(q, \tilde{\nu})
    = \frac{1}{N_\mb{k}} \sum_{\tilde{\mb{k}}} G(k) G(k + q)
    - g(\tilde{\nu}) g(\tilde{\nu} + \omega) \,.
\end{equation}

To use the SDE for $\Sigma$, we need $F^=$.
This full (nonlocal) vertex is obtained by adding the full vertices with ladder contributions in the $a$ and $t$ channels and then subtracting the local vertex to prevent double counting:
\begin{multline} \label{eq:lDGA_full_vertex_1}
    F^=(q, k, k')_t
    = (i_t^= + \Phi^=_t)(q, \nu, \nu')
    \\
    + (i_a^= + \Phi^=_a)(k'-k, \omega+\nu, \nu)
    - f^=(\omega, \nu, \nu')_t \,.
\end{multline}
Converting the $a$ and $t$ channels to the M and D channels using \Eqs{eq:physical_spin_F} and \eqref{eq:physical_spin_Phi_cross}, we find
\begin{align} \label{eq:lDGA_full_vertex}
    F^=(q, k, k')_t
    &= \frac{1}{2}(i_\Dch + \Phi_\Dch)(q, \nu, \nu')
    - \frac{1}{2}(i_\Mch + \Phi_\Mch)(q, \nu, \nu')
    \nnnl
    &- (i_\Mch + \Phi_\Mch)(k'-k, \omega+\nu, \nu)
    \nnnl
    &- \frac{1}{2} f_\Dch(\omega, \nu, \nu')_t
    + \frac{1}{2} f_\Mch(\omega, \nu, \nu')_t \,.
\end{align}
Note that the full vertex has momentum dependence on all three arguments, while the reducible vertex depends only on the bosonic momentum.

Inserting this full vertex into the SDE \eqref{eq:SDE}, we find
\begin{multline} \label{eq:lDGA_SDE_der1}
    \Sigma(k) - \Sigma_\Hart
    = \frac{1}{2} \sum_q G(k+q) \biggl[ 
        -\bigl( (i_\Dch + \Phi_\Dch) \Pi_t U \bigr)(q, k)
        \\
        + 3\bigl( (i_\Mch + \Phi_\Mch) \Pi_t U \bigr)(q, k)
        +  \bigl( (f_\Dch - f_\Mch) \Pi_t U \bigr)(q, k)
    \biggr] \,.
\end{multline}
The first and second terms in the square bracket are the $\nu \to \infty$ limit of \Eq{eq:lDGA_BSE} (in left-right mirrored form):
\begin{equation}
    \bigl[ (i_\MDch + \Phi_\MDch) \Pi_t U \bigr](q, \nu)
    = \mp K_{12,\MDch}(q, \nu) \,,
\end{equation}
where we defined $K_{12,\MDch} = K_{1,\MDch} + K_{2,\MDch}$.
The $\mp$ sign comes from the different sign of the bare vertex in the M and D channels [\Eq{eq:bare_vertex}].
The SDE then reads
\begin{multline} \label{eq:lDGA_SDE_K12}
    \Sigma(k) - \Sigma_\Hart
    = \frac{1}{2} \sum_q G(k+q) \biggl[ 
        -K_{12,\Dch}(q, \nu)
        \\
        - 3 K_{12,\Mch}(q, \nu)
        + \bigl( (f_\Dch - f_\Mch) \Pi_t U \bigr)(q, k)
    \biggr] \,.
\end{multline}

The $K_{12}$ asymptote thus suffices to evaluate the SDE, and we only compute this asymptote instead of the whole reducible vertex.
Taking $\nu' \to \infty$ in \Eq{eq:lDGA_fdBSE}, we get
\begin{multline} \label{eq:lDGA_fdBSE_K12}
    \tilde{K}_{12,\MDch}(q, \nu)
    = 
    \frac{1}{\beta} \sum_{\tilde{\nu}} f_{\MDch}(\omega, \nu, \tilde{\nu}) \,
    \tilde{\Pi}_t(q, \tilde{\nu})
    \\
    \times \bigl( K_{12,\MDch}(q, \tilde{\nu}) \mp U \bigr) \,.
\end{multline}
\EQ{eq:lDGA_fdBSE_K12} is numerically beneficial over \Eq{eq:lDGA_fdBSE} as the variable to solve for has only two arguments ($q, \nu$) instead of three ($q, \nu, \nu'$).
Further taking $\nu \to \infty$, we find
\begin{multline} \label{eq:lDGA_fdBSE_K1}
    \tilde{K}_{01,\MDch}(q)
    =
    \frac{1}{\beta} \sum_{\tilde{\nu}} (k_{1\bar{2},\MDch}(\omega, \tilde{\nu}) \mp U) \,
    \tilde{\Pi}_t(q, \tilde{\nu})
    \\
    \times \bigl( K_{12,\MDch}(q, \tilde{\nu}) \mp U \bigr) \,.
\end{multline}
We solve Eqs.~\eqref{eq:lDGA_fdBSE_K12} and \eqref{eq:lDGA_fdBSE_K1} to compute the asymptotes of the nonlocal vertex, and then compute the susceptibility and the Hedin vertex using Eqs.~\eqref{eq:lDGA_chi_and_hedin}.

Next, we apply the Moriyaesque $\lambda$ correction for the magnetic susceptibility~\cite{Katanin2009,Rohringer2011}:
\begin{equation}
    \chi^{\lambda_\Mori}_\Mch(q)
    = \frac{1}{1 / \chi_\Mch(q) + \lambda_\Mori} \,.
\end{equation}
The parameter $\lambda_\Mori$ is found by enforcing the equal-spin sum rule [\Eq{eq:sum_rule_chi_equal}]:
\begin{equation} \label{eq:ldga_sum_rule}
    \frac{1}{2} \sum_q \Bigl( \chi_\Dch(q) + \chi^{\rm \lambda_\Mori}_\Mch(q) \Bigr) = n_\sigma (1-n_\sigma) \,.
\end{equation}
The sum rule requires careful treatment of the high-frequency tail, which decays as $1/\omega^2$~\cite{Krien2017}.
With $\chi$ computed on Matsubara frequencies $i\omega_n = 2\pi i n/\beta$ for $n = -N, \dots, N$, we evaluate the frequency sum [\Eq{eq:ldga_sum_rule}] as
\begin{multline}
    \sum_{\omega} \chi(\omega)
    = \sum_{n = -N}^{N} \chi(i\omega_n)
    \\
    + \Bigl( \chi(i\omega_{-N}) + \chi(i\omega_N) \Bigr) \sum_{n = N+1}^{\infty} \frac{N^2}{n^2} \,.
\end{multline}
We also apply this scheme to $\chi$ from DMFT to converge the sum beyond the box size of the DMFT calculation.

The $\lambda$-corrected $\chi$ is used for a corrected $\Sigma$. With
\begin{equation}
    K_{12,\MDch}(q, \nu) = U [(U \chi_{\MDch}(q) \mp 1) \lambda_\MDch(q, \nu) \pm 1] \,,
\end{equation}
\begin{equation}
    K_{12,\MDch}(q, \nu) = \pm U [(1 \mp U \chi_{\MDch}(q)) \lambda_\MDch(q, \nu) - 1] \,,
\end{equation}
we rewrite \Eq{eq:lDGA_SDE_K12} in terms of $\chi$ and the Hedin vertex, 
\begin{align} \label{eq:lDGA_SDE_hedin}
    &\Sigma(k) - \Sigma_\Hart
    = \frac{U}{2} \sum_q G(k+q) \biggl[ 
        \bigl( U \chi_\Dch(q) + 1 \bigr) \lambda_\Dch(q, \nu) - 1
        \nnnl
        &\quad + 3 \bigl[ (U \chi_\Mch(q) - 1) \lambda_\Mch(q, \nu) + 1 \bigr]
        \nnnl
        &\quad +\frac{1}{\beta} \sum_{\tilde{\nu}} \Pi_t(q, \tilde{\nu})
        \bigl( f_\Dch(\omega, \nu, \tilde{\nu}) - f_\Mch(\omega, \nu, \tilde{\nu}) \bigr)
    \biggr] \,,
\end{align}
where
\begin{equation}
    \Pi_t(q, \tilde{\nu})
    = \frac{1}{N_\mb{k}} \sum_{\tilde{\mb{k}}} G(\tilde{k}) G(\tilde{k}+q) \,.
\end{equation}
This result is equivalent to Eq.~(8) of Ref.~\cite{Katanin2009}, except for sign differences due to the sign of the bare vertex.

\subsection{Preconditioning} \label{sec:preconditioning}
\subsubsection{Preconditioned fixed-point iteration}
A preconditioner is a function that approximates the problem to be solved but is simple enough so that its inverse is easy to compute.
Let us consider a vector-valued self-consistent problem $x = \mc{G}(x)$.
The naive fixed-point iteration reads
\begin{subequations} \label{eq:prec_simple_naive}
\begin{align}
    \label{eq:prec_simple_naive_1}
    y^{(i)} &= \mc{G}(x^{(i)}) - x^{(i)} \,,
    \\
    \label{eq:prec_simple_naive_2}
    x^{(i+1)} &= x^{(i)} + y^{(i)} = \mc{G}(x^{(i)}) \,.
\end{align}
\end{subequations}
Here, one computes the residual $y^{(i)}$ and adds it to the current guess $x^{(i)}$ to update the guess.

Now, let us consider the linearized form of $\mc{G}$ around $x=0$,
\begin{equation}
    \mc{G}(x) = g_0 + Gx + O(x^2) \,.
\end{equation}
Here, $Gx$ denotes a matrix-vector multiplication.
We now use $\one - G$ as a preconditioner for the fixed-point iteration, modifying the iteration as
\begin{subequations} \label{eq:prec_simple_prec}
\begin{align}
    \label{eq:prec_simple_prec_1}
    y^{(i)} &= \mc{G}(x^{(i)}) - x^{(i)} \,,
    \\
    \label{eq:prec_simple_prec_2}
    z^{(i)} &= (\one - G)^{-1} y^{(i)} \,,
    \\
    \label{eq:prec_simple_prec_3}
    x^{(i+1)} &= x^{(i)} + z^{(i)} \,.
\end{align}
\end{subequations}
The difference with the naive iteration \Eq{eq:prec_simple_naive} is that we multiply $(\one - G)^{-1}$ to the residual $y^{(i)}$ before adding it to the guess $x^{(i)}$.
We note that an explicit computation of the matrix inverse $(\one - G)^{-1}$ can be avoided by using iterative linear solvers such as the generalized minimal residual (GMRES) method.

If $\mc{G}$ is a linear function, i.e., $\mc{G}(x) = g_0 + Gx$, a single preconditioned iteration starting from $x^{(0)} = 0$ yields the exact solution $x^* = (\one - G)^{-1} g_0$:
\begin{equation}
\begin{gathered}
    x^{(0)} = 0,\ y^{(0)} = \mc{G}(0) = g_0 \,,
    \\
    x^{(1)} = z^{(0)} = (\one - G)^{-1} g_0 = x^* \,.
\end{gathered}
\end{equation}
If $\mc{G}$ is a nonlinear function, one needs multiple iterations to converge to the exact solution.
Still, preconditioning can help speed up and stabilize the iterative procedure.
Also, we note that the preconditioned iteration \Eq{eq:prec_simple_prec} can be combined with convergence acceleration methods such as the Anderson acceleration.

\subsubsection{Preconditioned fd-\DGA iteration}
The fd-\DGA equation is a nonlinear equation for the fd vertices.
Due to the nonlinearity, simply iterating the fd-BSE often leads to slow convergence, instability, and even divergence.
Here, we discuss a way to mitigate this problem using preconditioning.

We begin with repeating the fd-\DGA equation \eqref{eq:fdBSE} for convenience:
\begin{equation}
\tilde{\Phi}_r
= f \tilde{\Pi}_r F + \tilde{I}_r \Pi_r F + f \pi_r \tilde{I}_r \Pi_r F + f \pi_r \tilde{I}_r
.
\nonumber
\end{equation}
With the starting point
\begin{equation} \label{eq:precond_initial}
    \tilde{\Phi}_r^{(0)} = 0 \,,
    \ \tilde{I}_r^{(0)} = 0 \,,
    \ F^{(0)} = f + \tilde{\Phi}_r^{(0)} + \tilde{I}_r^{(0)} = f \,,
\end{equation}
and iterating by inserting the vertices of the previous iterations to the right of \Eq{eq:fdBSE}, we find
\begin{subequations}
\begin{align}
    \tilde{\Phi}_r^{(1)}
    &= f \tilde{\Pi}_r f \,,
    \\
    \tilde{\Phi}_r^{(2)}
    &= f \tilde{\Pi}_r (f + \tilde{F}^{(1)}) + (\one + f \pi_r) \tilde{I}_r^{(1)} \Pi_r (f + \tilde{F}^{(1)})
    \nnnl
    &+ f \pi_r \tilde{I}_r^{(1)} \,.
\end{align}
\end{subequations}
This series does not easily converge even when $\tilde{\Pi}$ is small, because each iteration multiplies terms like $f \pi_r$ to the vertex of the previous iteration.
Since $f \pi_r$ is given by the starting point of the fd-\DGA problem, it cannot be controlled by having a small difference $\tilde{\Pi}$.
We were not able to converge this fd-\DGA iteration for the HM with the parameters of \Fig{fig:Wu_self_energy}, even with the help of convergence acceleration methods.

To solve this problem, we precondition the fixed-point iteration as in \Eq{eq:prec_simple_prec}.
We write $x^{(i)}$, $y^{(i)}$, and $z^{(i)}$ as $\tilde{\Phi}_r^{(i)}$, $\tilde{\Psi}_r^{(i)}$, and $\tilde{\Xi}_r^{(i)}$, respectively.
All quantities share the same frequency and momentum dependence of the reducible vertex  $\Phi_r$.
Note that the equations for the three channels $r=a,p,t$ are coupled and should be solved simultaneously.
We begin with an initial guess for the finite difference vertex.
Starting from the initial condition of \Eq{eq:precond_initial} at $i=0$, at the $i$-th iteration, we first compute the residual of the self-consistent fd-\DGA equation:
\begin{subequations}  \label{eq:prec_fddga_prec}
\begin{equation} \label{eq:prec_fddga_prec_1}
    \tilde{\Psi}^{(i)}_r
    = f \tilde{\Pi}_r F^{(i)}
    + (\one + f \pi_r) \tilde{I}_r^{(i)} \Pi_r F^{(i)}
    + f \pi_r \tilde{I}_r^{(i)}
    - \tilde{\Phi}^{(i)}_r \,.
\end{equation}

Now, we precondition the residual by using the linearized fd-parquet equation \eqref{eq:fdBSE}.
Linearization at the reference vertex is achieved by replacing $\Pi_r$ and $F$ in \Eq{eq:fdBSE} with the reference quantities $\pi_r$ and $f$:
\begin{multline}
    \underbrace{\tilde{\Psi}_r}_{\mc{G}(x)}
    = \underbrace{f \tilde{\Pi}_r f}_{g_0}
    + \underbrace{(\one + f \pi_r) \tilde{I}_r (\one + \pi_r f)
    - \tilde{I}_r}_{Gx}
    \\
    + O(\tilde{\Pi}, \tilde{I})^2 \,.
\end{multline}
The preconditioned residual $z^{(i)}$ in \Eq{eq:prec_simple_prec_2} can be obtained by solving a \emph{linear} equation $z^{(i)} = y^{(i)} + G z^{(i)}$.
For the fd-\DGA, the equation reads
\begin{equation} \label{eq:prec_fddga_prec_2}
    \underbrace{\tilde{\Xi}_r^{(i)}}_{z^{(i)}}
    = \underbrace{\tilde{\Psi}^{(i)}_r}_{y^{(i)}}
    + \underbrace{(\one + f \pi_r) \sum_{s \neq r} \tilde{\Xi}_{s}^{(i)} (\one + \pi_r f)
    - \sum_{s \neq r} \tilde{\Xi}_{s}^{(i)}}_{G z^{(i)}} \,.
\end{equation}
This equation generalizes the multiloop functional renormalization group (mfRG) equation with $\tilde{\Psi}^{(i)}_r$ replacing the one-loop term.
Note that the bubbles and vertices multiplied to the left and right of the irreducible vertex $\tilde{\Xi}_{s}^{(i)}$ are the reference ones, $\pi_r$ and $f$.
Equation~\eqref{eq:prec_fddga_prec_2} is easier to solve than the original fd-parquet equation \eqref{eq:fdBSE} thanks to its linearity.
In practice, we solve it using the GMRES method.

Finally, we update the fd vertices as
\begin{equation} \label{eq:prec_fddga_prec_3}
\begin{gathered}
    \tilde{\Phi}_r^{(i+1)}
    = \tilde{\Phi}_r^{(i)}
    + \tilde{\Xi}_r^{(i)} \,,
    \\
    \tilde{I}_r^{(i+1)} = {\textstyle \sum_{s \neq r}} \tilde{\Phi}_{s}^{(i+1)} \,,
    \\
    \tilde{F}^{(i+1)} = \tilde{\Phi}_r^{(i+1)} + \tilde{I}_r^{(i+1)} \,.
\end{gathered}
\end{equation}
\end{subequations}
Equations~\eqref{eq:prec_fddga_prec_1}--\eqref{eq:prec_fddga_prec_3} follow the structure of Eqs.~\eqref{eq:prec_simple_prec_1}--\eqref{eq:prec_simple_prec_3}.
We solve the fixed-point problem formed by Eqs.~\eqref{eq:prec_fddga_prec_1}--\eqref{eq:prec_fddga_prec_3} using the restarted Anderson acceleration.

\begin{table*} 
\centering
\begin{tabular}{c||l||l|l||l|l||l|l||}
&
\multicolumn{1}{c||}{SIAM, fd-PA ($\beta=20$)} &
\multicolumn{2}{c||}{PHS HM, fd-p\DGA (\Fig{fig:highT})} &
\multicolumn{2}{c||}{Doped HM, fd-p\DGA (\Fig{fig:Wu_self_energy})} &
\multicolumn{2}{c||}{Doped HM, $\lambda$-\lDGA (\Fig{fig:nsc_ladder})}
\\ \cline{2-8}
&
\multicolumn{1}{c||}{\# frequencies} &
\multicolumn{1}{c|}{\# frequencies} & \multicolumn{1}{c||}{\# momentum} &
\multicolumn{1}{c|}{\# frequencies} & \multicolumn{1}{c||}{\# momentum} &
\multicolumn{1}{c|}{\# frequencies} & \multicolumn{1}{c||}{\# momentum}
\\ \hline
$G$           & 4000 & 40 & $48 \times 48$ & 112 & $48 \times 48$ & 127 & $160 \times 160$ \\
$\Sigma$      & 600  & 40 & $48 \times 48$ & 112 & $48 \times 48$ & 127 & $160 \times 160$ \\
$\tilde{K}_1$ & 1501 & 37 & $4  \times 4$  & 111 & $15 \times 15$ & 127 & $160 \times 160$ \\
$\tilde{K}_2$ & $801 \times 700$
& $9 \times 10$ & $4 \times 4$
& $27 \times 28$ & $15 \times 15$
& $127 \times 128$ & $160 \times 160$ \\
$\tilde{K}_3$ & $701 \times 600 \times 600$
& $9 \times 10 \times 10$ & $4 \times 4$
& $27 \times 28 \times 28$ & $15 \times 15$
& $127 \times 128 \times 128$ & $160 \times 160$
\end{tabular}
\caption{Frequency and momentum parameters for the fd-PA, fd-p\DGA, and $\lambda$-\lDGA calculations.
}
\label{tab:parameters}
\end{table*}

\subsubsection{Analysis of the preconditioned fd-\DGA iteration}
Below, we show that the $i$-th residual of the preconditioned fixed-point iteration is of order $O(\tilde{\Pi}^{i+1})$.
This dependence contrasts with the $O(\Pi^{i+1})$ scaling of the naive fixed-point iteration, in that the residual depends on the fd bubble instead of the full one.
Since the magnitude of the fd bubble is typically smaller than the full bubble, the preconditioned fd-\DGA iteration is more stable than the naive one.

Let us begin with the initial iteration.
As we start from a zero initial guess [\Eq{eq:precond_initial}], the initial residual is the 1-loop term:
\begin{equation}
    \tilde{\Psi}^{(0)} = f \tilde{\Pi}_r f \,.
\end{equation}
The preconditioned residual is the solution of the mfRG equation:
\begin{equation} \label{eq:prec_fddga_mfRG_iter1}
    \tilde{\Xi}_r^{(0)}
    =f \tilde{\Pi}_r f
    + (\one + f \pi_r) \sum_{s \neq r} \tilde{\Xi}_{s}^{(0)} (\one + \pi_r f)
    - \sum_{s \neq r} \tilde{\Xi}_{s}^{(0)} \,.
\end{equation}
The updated vertex is $\tilde{\Phi}_r^{(1)} = \tilde{\Xi}_r^{(0)}$.
This vertex agrees with the exact solution to linear order in $\tilde{\Pi}$, as the mfRG is a linearized form of the fd-\DGA equations.
In other words, the residual at $i=1$ is at least quadratic in $\tilde{\Pi}$.
This fact can be proved explicitly:
\begin{align}
    &\tilde{\Psi}^{(1)}
    = f \tilde{\Pi}_r F^{(1)}
    + (\one + f \pi_r) \tilde{I}_r^{(1)} \Pi_r F^{(1)}
    + f \pi_r \tilde{I}_r^{(1)}
    - \tilde{\Phi}^{(1)}_r
    \nnnl
    &= f \tilde{\Pi}_r f
    + (\one + f \pi_r) \tilde{I}_r^{(1)} \pi_r f
    + f \pi_r \tilde{I}_r^{(1)}
    - \tilde{\Phi}^{(1)}_r
    + O(\tilde{\Pi}^2)
    \nnnl
    &= 0 + O(\tilde{\Pi}^2) \,.
\end{align}
In the second equality, we used
\begin{equation}
    \tilde{\Phi}^{(1)}_r,\ \tilde{I}^{(1)}_r,\ \tilde{F}^{(1)} = O(\tilde{\Pi})
\end{equation}
to collect only the terms zeroth- and first-order in $\tilde{\Pi}$.
In the third equality, we used the mfRG equation \eqref{eq:prec_fddga_mfRG_iter1} of the previous iteration to find that the linear-order terms add up to zero.
Note that we also used
\begin{equation}
    \tilde{\Phi}^{(1)} = \tilde{\Xi}^{(0)},\quad
    \tilde{I}^{(1)} = \sum_{s \neq r} \tilde{\Xi}_{s}^{(0)} \,.
\end{equation}

Now, we use mathematical induction to show that the residual at the $i$-th iteration is of order $O(\tilde{\Pi}^{i+1})$.
We assume at the $(i-1)$-th iteration we have
\begin{equation} \label{eq:prec_induction_assumption}
    \tilde{\Psi}^{(i-1)},\ \tilde{\Xi}^{(i-1)},\ \tilde{\Phi}^{(i)} - \tilde{\Phi}^{(i-1)}
    = O(\tilde{\Pi}^{i}) \,.
\end{equation}
By subtracting \Eq{eq:prec_fddga_prec_1} for $i$-th and $(i-1)$-th iterations, we find that the $i$-th residual is
\begin{align} \label{eq:prec_induction_1}
    \tilde{\Psi}^{(i)}_r
    &= \tilde{\Psi}^{(i-1)}_r + f \tilde{\Pi}_r (\tilde{F}^{(i)} - \tilde{F}^{(i-1)})
    \nnnl
    &\quad + (\one + f \pi_r) (\tilde{I}_r^{(i)} \Pi_r F^{(i)} - \tilde{I}_r^{(i-1)} \Pi_r F^{(i-1)})
    \nnnl
    &\quad + f \pi_r (\tilde{I}_r^{(i)} - \tilde{I}_r^{(i-1)})
    \nnnl
    &\quad - (\tilde{\Phi}^{(i)}_r - \tilde{\Phi}^{(i-1)}_r) \,.
\end{align}
Using the inductive assumption [\Eq{eq:prec_induction_assumption}] to collect the $O(\tilde{\Pi}^{i+1})$ terms, we find
\begin{align} \label{eq:prec_induction_2}
    \tilde{\Psi}^{(i)}_r
    &= \tilde{\Psi}^{(i-1)}_r
    + (\one + f \pi_r) (\tilde{I}_r^{(i)} - \tilde{I}_r^{(i-1)}) \pi_r f
    \nnnl
    &\quad + f \pi_r (\tilde{I}_r^{(i)} - \tilde{I}_r^{(i-1)})
    - \tilde{\Xi}^{(i-1)}_r
    + O(\tilde{\Pi}^{i+1})
    \nnnl
    &= \tilde{\Psi}^{(i-1)}_r
    + (\one + f \pi_r) \sum_{s \neq r} \tilde{\Xi}^{(i-1)}_{s} (1 + \pi_r f)
    - \sum_{s \neq r} \tilde{\Xi}^{(i-1)}_{s}
    \nnnl
    &\quad - \tilde{\Xi}^{(i-1)}_r
    + O(\tilde{\Pi}^{i+1}) \,.
\end{align}
In the second equality, we used
\begin{equation}
    \tilde{I}_r^{(i)} - \tilde{I}_r^{(i-1)}
    = \sum_{s \neq r} \bigl( \tilde{\Phi}_{s}^{(i)} - \tilde{\Phi}_{s}^{(i-1)} \bigr)
    = \sum_{s \neq r} \tilde{\Xi}^{(i-1)}_{s} \,,
\end{equation}
which follows from the first and second relations of \Eq{eq:prec_fddga_prec_3}.
The third line of \Eq{eq:prec_induction_2} is identical to the right side of the preconditioning equation \Eq{eq:prec_fddga_prec_2} at the $(i-1)$-th iteration.
Therefore, we find
\begin{equation} \label{eq:prec_induction_3}
    \tilde{\Psi}^{(i)}_r
    = \tilde{\Xi}_r^{(i)} - \tilde{\Xi}^{(i-1)}_r + O(\tilde{\Pi}^{i+1})
    = O(\tilde{\Pi}^{i+1}) \,.
\end{equation}
Since the preconditioned residual $\tilde{\Xi}^{(i)}_r$ depends linearly on $\tilde{\Psi}^{(i)}_r$ with the linear relation \Eq{eq:prec_fddga_prec_2} that does not explicitly depend on $\tilde{\Pi}$, its $\tilde{\Pi}$-scaling should be the same as that of $\tilde{\Psi}^{(i)}_r$:
\begin{equation}
    \tilde{\Xi}^{(i)}_r
    = O(\tilde{\Pi}^{i+1}) \,.
\end{equation}
Finally, by the update rule \Eq{eq:prec_fddga_prec_3}, we find
\begin{equation}
    \tilde{\Phi}_r^{(i+1)}
    - \tilde{\Phi}_r^{(i)}
    = \tilde{\Xi}_r^{(i)}
    = O(\tilde{\Pi}^{i+1}) \,.
\end{equation}
By mathematical induction, this concludes the proof that \Eq{eq:prec_induction_assumption} holds for all $i$.

\subsection{Numerical implementation}

We implement the parquet equations for the AIM and the Hubbard model using the \texttt{MatsubaraFunctions.jl} software library~\cite{Kiese2024a,Kiese2024b}.
To speed up the calculation and avoid unphysical solutions, we enforce the crossing, time-reversal, complex conjugation, and reciprocal space crystal symmetries using the respective interface of the library.

We make sure that the calculations are sufficiently accurate with respect to frequency box sizes.
An overview of the number of frequencies per asymptotic class and the size of the bosonic momentum mesh is shown in Table~\ref{tab:parameters}.

For the analytic continuation of the self-energy in \Fig{fig:Wu_spectral_3d}(b), we used the \textsc{ana\_cont} package~\cite{Kaufmann2023anacont}.
Since the propagation of stochastic noise from the quantum Monte Carlo impurity solver to the p\DGA self-energy is not easy to calculate, we set the noise amplitude to 0.005, which was the smallest value where the analytic continuation process did not fail.
We set the preblur parameter to 1.0.

\begin{table*}[]
\begin{tabular}{c|c|c|c|c|c|c||c|c}
 & \multicolumn{2}{c|}{$n_\sigma (1 - n_\sigma)$}
 & \multicolumn{2}{c|}{$\langle n_\sigma n_{\bar{\sigma}} \rangle$}
 & \multicolumn{2}{c||}{$-E_{\rm kin}$}
 & \multirow{2}{*}{$\sum\chi_\Mch$}
 & \multirow{2}{*}{$\sum\chi_\Dch$}
\\
 \cline{1-7}
 & $\sum\chi_{\sigma\sigma}$  
 & $-\!{\displaystyle \lim_{\nu \to \infty}}\! \nu \frac{\im\Sigma(\nu)}{U^2}$
 & $\sum\chi_{\sigma\bar{\sigma}} \!+\! n_\sigma^2$
 & $\frac{1}{U} \sum G \Sigma$ 
 & ${\displaystyle \lim_{\omega \to \infty}} (i\omega)^2 \chi_{\rm M,D}$
 & $-\! \sum \! \varepsilon_\mb{k} \langle n_{\mb{k}\sigma} \rangle$ & & \\ \hline
Exact (DiagMC) & \multicolumn{2}{c |}{0.2496} &  \multicolumn{2}{c |}{0.0755(5)} & \multicolumn{2}{c ||}{1.112(6)} & 0.4044(5) & 0.0948(5) \\ \hline
DMFT
& 0.250 & 0.25 & 0.094 & 0.094 & 1.2, 1.2 & 1.17 & 0.386 & 0.114 \\
p\DGA
& 0.267 & 0.30 & 0.091 & 0.058 & 1.8, 1.8 & 1.11 & 0.406 & 0.128 \\
p\DGA+loc.~$\Sigma$
& 0.267 & 0.25 & 0.091 & 0.075 & 1.8, 1.8 & 1.16 & 0.406 & 0.128 \\
p\DGA+$K_1$-post
& 0.248 & 0.24 & 0.060 & 0.070 & 1.8, 1.0 & 1.12 & 0.418 & 0.078
\end{tabular}
\caption{Sum rules~\eqref{eq:sum_rule_chi}, \eqref{eq:sum_rule_Sigma}, and \eqref{eq:sum_rule_kinetic}, for the HM at parameters chosen for \Fig{fig:Wu_self_energy}, where the occupation $n_\sigma = 0.48$.
The susceptibilities in the headers denote local ($\mathbf{q}$-averaged) quantities.
The exact result is derived using $\langle n_\sigma n_{\bar{\sigma}} \rangle = 0.0755(5)$ and $E_{\rm kin} = -1.112(6)$ obtained from DiagMC. We have implemented diagrammatic Monte Carlo in its connected determinant form~\cite{Rossi2017} with many-configuration sampling~\cite{Simkovic2021} using the TRIQS library~\cite{Parcollet2015}. 
The local $\Sigma$ correction (loc.~$\Sigma$) uses \Eq{eq:local_corr} for the self-energy, while the susceptibility remains the same.
The $K_1$-postprocessing correction ($K_1$-post) uses \Eq{eq:tail_K1_pp} for the susceptibility and \Eq{eq:tail_sigma_corr} for the self-energy.
For the $\omega \to \infty$ limit of the susceptibility, we evaluate the susceptibility at Matsubara frequencies $\omega = [30, \dots, 50]  \tfrac{2\pi}{\beta}$ and show the average.
See \Fig{fig:Wu_sum_rule} for the convergence of the sum rules and the high-frequency limites with respect to the maximum frequency.
}
\label{tab:sum_rule}
\end{table*}

\section{Additional results}

\subsection{$\lambda$-corrected \lDGA and PA}
\begin{figure}[tb]
\centering
\includegraphics[width = 0.99\linewidth]{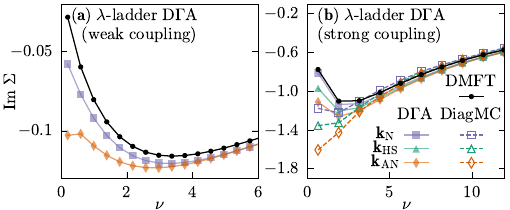}
\caption{
Frequency dependence of the self-energy from $\lambda$-\lDGA with Moriyaesque $\lambda_\Mori$ correction in the magnetic channel.
(a) Weak-coupling HM at PHS with $U=2.0$, $T=0.063$, $t' = 0$, where $\lambda_\Mori = 0.147$.
Node and antinode are located at $\kN=(\pi/2, \pi/2)$ and $\kAN = (\pi, 0)$.
The hot spot is not specified since the HM at PHS with $t'=0$ has perfect nesting and every point on the Fermi surface is a hot spot.
(b) Hole-doped HM with parameters as in \Fig{fig:Wu_self_energy}, where $\lambda_\Mori = 0.244$.
}
\label{fig:nsc_ladder}
\end{figure}

In the main text, we used the self-consistent \lDGA~\cite{Kaufmann2021}, where the self-energy is updated (via the SDE) together with the vertex.
Alternatively, one may use the (non-self-consistent) \lDGA with the Moriyaesque $\lambda$ correction ($\lambda$-\lDGA)~\cite{Katanin2009}---this approach was shown to feature a weak-coupling PG in the HM at PHS ($U=2$, $t'=0$)~\cite{Schaefer2021}.
Figure~\ref{fig:nsc_ladder}(a) shows the self-energy at $T=0.063$, where the antinode is insulating-like and the node metallic-like, in agreement with Ref.~\cite{Schaefer2021}.
Our onset temperature for the PG, $T_* \approx 0.063$, is similar to but slightly lower than that of Ref.~\cite{Schaefer2021} ($T_* = 0.065$).
This discrepancy could be due to different impurity solvers used (CT-INT quantum Monte Carlo here, while exact diagonalization with four bath sites in Ref.~\cite{Schaefer2021})
or different treatments of the high-frequency part of the vertex
(we used the asymptotic decomposition~\cite{Li2016, Wentzell2020}, while Ref.~\cite{Schaefer2021} did not).

Contrary to the weak-coupling case, $\lambda$-\lDGA cannot describe the (strong-coupling) PG for the HM at the parameters used in \Fig{fig:Wu_self_energy} of the main text (stronger interaction, next-nearest-neighbor hopping, and hole doping).
Indeed, Fig.~\ref{fig:nsc_ladder}(b) shows the self-energy computed using $\lambda$-\lDGA, which is similar to the result of self-consistent \lDGA~[\Fig{fig:Wu_self_energy}(c)] and does not produce an insulating antinode.

\begin{figure}[tb]
    \centering
    \includegraphics[width = 0.99\linewidth]{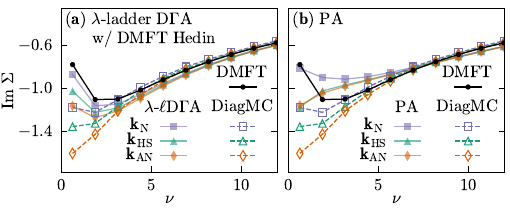}
    \caption{
    (a) Same as \Fig{fig:nsc_ladder}(b), but with the nonlocal correction to the Hedin vertex from \lDGA neglected.
    We find this approximation has little effect on the self-energy.
    (b) Same as \Fig{fig:nsc_ladder}(b), but using the PA instead of the \DGA.
    }
    \label{fig:Wu_PA_and_trilex}
\end{figure}

In \Fig{fig:Wu_PA_and_trilex}(a), the self-energy is computed with $\lambda$-\lDGA as in \Fig{fig:nsc_ladder}(b) but with the Hedin vertex taken from the DMFT impurity vertex.
We find that this additional approximation has little effect on the self-energy.
Thus, the effect of the \lDGA renormalization of the Hedin vertex on the PG is negligible [contrary to p\DGA, where it is crucial for the opening of the PG, see \Fig{fig:Wu_diagnostics}(b)].

Finally, we come back to the PA, which was mentioned in the conclusion of the main text.
Figure~\ref{fig:Wu_PA_and_trilex}(b) shows the self-energy computed in the PA.
The PA fails to reproduce the hierarchy of the antinode, hot spot, and node, overestimating the insulating behavior at the hot spot.
We attribute this to the long-ranged magnetic susceptibility in the PA and the overestimation of scattering with $(\pi, \pi)$ momentum transfer, see \Sec{sec:momentum_differentiation}~\cite{supplemental}.
Indeed, the PA captures the perturbative weak-coupling PG due to long-range spin fluctuation~\cite{Vilk1997, Schaefer2021} (and therein includes a perturbative feedback between the nonlocal channels not contained in the \lDGA),
but it does not capture the nonperturbative enhancement of the Hedin vertex (which actually involves antiferromagnetic spin fluctuations of relatively short length) needed for the strong-coupling PG.

\subsection{Sum rules and high-frequency limits}

\begin{figure*}[tb]
\centering
\includegraphics[width = 0.99\linewidth]{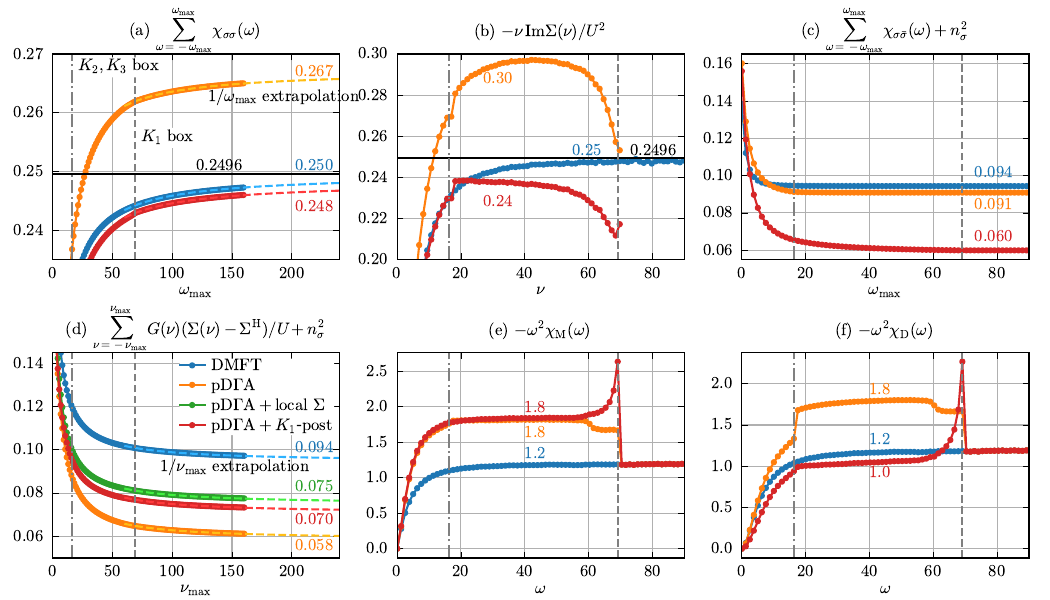}
\caption{
Convergence of the sum rules and asymptotic limits with respect to the maximum frequency included in the sum or the limit.
(a) \Eq{eq:sum_rule_chi_equal},
(b) \Eq{eq:sum_rule_Sigma_same},
(c) \Eq{eq:sum_rule_chi_diff},
(d) \Eq{eq:sum_rule_Sigma_diff}, and
(e, f) \Eq{eq:sum_rule_kinetic_a}.
The dot-dashed (dashed) vertical lines indicate the boundary of the frequency box for the $K_2$ and $K_3$ vertices ($K_1$ vertex).
For (a) and (d), we evaluate the sum using $1/\omega$ and $1/\nu$ extrapolation of the cumulative sum.
Since the infinite-frequency limits involving frequency multiplication in (b), (e), and (f) depend significantly on the box size, we take the value from the plateau region.
The p\DGA+local $\Sigma$ correction result (green) is shown only in (d), as it is identical to the p\DGA result for (a, c, e, f), and to the DMFT result for (b).
}
\label{fig:Wu_sum_rule}
\end{figure*}

We analyze the self-energy and susceptibilities by assessing the fulfillment of high-frequency limits and sum rules~\cite{Vilk1997, Krien2017}. Note that our definition of $\chi_{\MDch}$ [\Eq{eq:chi_def1}] is half of Eq.~(28) of Ref.~\cite{Vilk1997} and Eq.~(8) of Ref.~\cite{Krien2017}. The total electron density is $n = n_\sigma + n_{\bar{\sigma}} = 2n_\sigma$.
The sum rules for the susceptibilities in the physical channels read
[cf.~Eqs.~(35), (36) of Ref.~\cite{Vilk1997}]
\begin{subequations} \label{eq:sum_rule_chi}
\begin{align}
    \sum_q \chi_\Mch(q)
    &= n_\sigma - \langle n_\sigma n_{\bar{\sigma}} \rangle \,,
    \\
    \sum_q \chi_\Dch(q)
    &= \langle n_\sigma n_{\bar{\sigma}} \rangle + n_\sigma - 2 n_\sigma^2 \,.
\end{align}
The sum rules for the equal- and different-spin susceptibilities [\Eq{eq:chi_def1}] are
[cf.~Eqs.~(39), (45) of Ref.~\cite{Vilk1997}]
\begin{align}
    \label{eq:sum_rule_chi_equal}
    \sum_q \chi_{\sigma \sigma}(q)
    &= n_\sigma (1-n_\sigma) \,,
    \\
    \label{eq:sum_rule_chi_diff}
    \sum_q \chi_{\sigma \bar{\sigma}}(q) + n_\sigma^2
    &= \langle n_\sigma n_{\bar{\sigma}} \rangle
    = E_{\rm pot} / U \,.
\end{align}
\end{subequations}
The right sides of \Eq{eq:sum_rule_chi} also satisfy the following relations with the single-particle self-energy and Green's function
[cf.~Eqs.~(A9), (44) of Ref.~\cite{Vilk1997}]:
\begin{subequations} \label{eq:sum_rule_Sigma}
\begin{align}
    \label{eq:sum_rule_Sigma_same}
    - \frac{1}{U^2} \lim_{\nu \to \infty} \nu \im \Sigma(k) &= n_\sigma (1-n_\sigma) \,,
    \\
    \label{eq:sum_rule_Sigma_diff}
    \frac{1}{U} \sum_{k} G(k) \Sigma(k) &= \langle n_\sigma n_{\bar{\sigma}} \rangle \,.
\end{align}
\end{subequations}
Finally, the high-frequency tail of the susceptibilities relates to the single-particle occupation via the kinetic energy
[cf.~Eq.~(10) of Ref.~\cite{Krien2017}]:
\begin{subequations} \label{eq:sum_rule_kinetic}
\begin{align}
    \label{eq:sum_rule_kinetic_a}
    E_{\rm kin}
    &= - \lim_{\omega \to \infty} \frac{1}{N_q} \sum_\mb{q} (i\omega)^2 \chi_{\MDch}(\omega, \mb{q})
    \\
    \label{eq:sum_rule_kinetic_b}
    &= \frac{1}{N_\mb{k}} \sum_{\mb{k} \sigma} \varepsilon_\mb{k} \langle n_{\mb{k} \sigma} \rangle \,.
\end{align}
\end{subequations}

We evaluate these relation with the self-energy and vertices from p\DGA and summarize the results in Table~\ref{tab:sum_rule}. The convergence of the infinite sum and the high-frequency limit is shown in \Fig{fig:Wu_sum_rule}. For the infinite sum, we use $1/\omega$ extrapolation [\Figs{fig:Wu_sum_rule}(a, d)]. For high-frequency limits involving multiplication by $\nu$ or $\omega^2$, we take the value at the plateau region [\Figs{fig:Wu_sum_rule}(b, e, f)]. Note that \Eq{eq:sum_rule_kinetic_a}, relating the kinetic energy to the high-energy tail of the susceptibility is particularly strongly violated in p\DGA. With the $K_1$-postprocessing correction, the M-channel contribution to the D and S channels is significantly damped, improving the agreement, but this is likely a coincidence.

\subsection{Momentum differentiation of the self-energy} \label{sec:momentum_differentiation}

\begin{figure*}[tb]
\centering
\includegraphics[width = 0.99\linewidth]{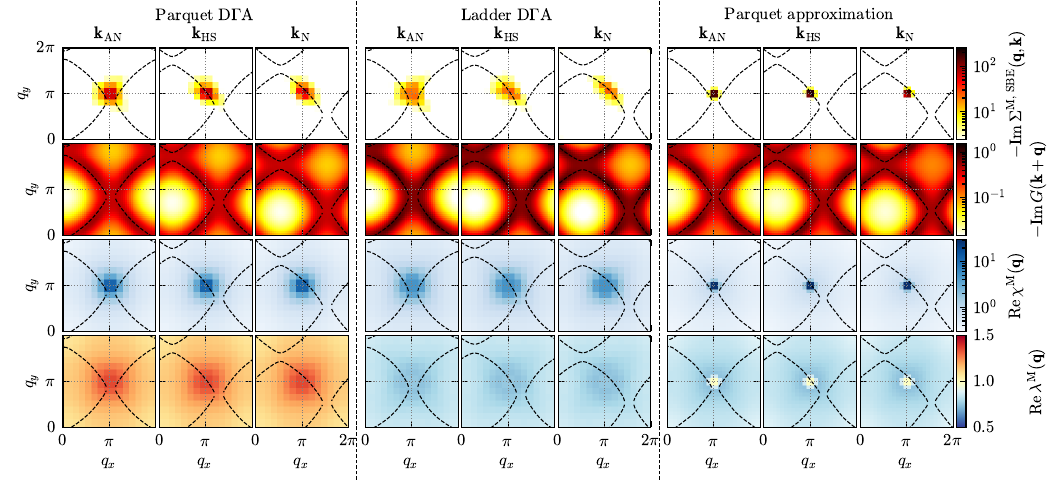}
\caption{Momentum-space diagnostics of $\Sigma$ [\Eq{eq:q_decomposition}] at the lowest Matsubara frequency $\nu = i\pi T$ for the HM, parameters as in \Fig{fig:Wu_self_energy}. Rows 2 to 4 show the three factors in the summand of \Eq{eq:q_decomposition} at $\omega = 0$. Dashed curves indicate the Fermi surface.}
\label{fig:Wu_diagnostics_in_q}
\end{figure*}

An important signature of the strong-coupling PG is that the single-particle properties are most strongly insulating at the AN, less insulating at the HS, and metallic at the N~\cite{Wu2017, Simkovic2024}.
This hierarchy differs from the weak-coupling PG, where the gap opening is strongest at the HS due to Fermi-surface nesting with $(\pi, \pi)$ antiferromagnetic fluctuations.
(The differentiation of HS from N and AN is not present in the $t'=0$ HM at PHS due to the perfect nesting of all $\mb{k}$ points on the Fermi surface.)
Here, we analyze the origin of the momentum-space differentiation in the self-energy and the aforementioned hierarchy.

As shown in \Fig{fig:Wu_diagnostics}, magnetic fluctuations drive the PG and the momentum differentiation. Here, we further decompose this magnetic contribution in momentum space using the SBE representation of the SDE, where the summand is the product of the Green's function, the susceptibility, and the Hedin vertex~\cite{Yu2024a}:
\begin{equation} \label{eq:q_decomposition}
    \Sigma^{\sbe}_{\mb{q},\,\Mch}(k)
    = \frac{3U^2}{2 \beta} \sum_{\omega} G(k+q)
    \chi_\Mch(q) \lambda_\Mch(q, \nu) \,.
\end{equation}
Due to the $s$-wave approximation of the vertex, the Hedin vertex does not depend on the fermionic momentum $\mb{k}$.

In the first row of \Fig{fig:Wu_diagnostics_in_q}, we show the decomposition \Eq{eq:q_decomposition} at the lowest Matsubara frequency $\nu = i\pi T$ and at three momentum points, AN, HS, and N.
In the second to fourth rows, we show the three factors in the summand of \Eq{eq:q_decomposition} for their $\omega = 0$ slices, which are the largest in magnitude.
The product of these three rows mostly explains the self-energy in the first row.

We now analyze whether the peak of the self-energy contribution is located at the commensurate transfer momentum $\mb{q} = (\pi, \pi)$ or shifted to an incommensurate $\mb{q} \neq (\pi, \pi)$.
For p\DGA, the self-energy diagnostics for the AN is peaked at $\mb{q} = (\pi, \pi)$, but it is shifted to an incommensurate momentum around $\mb{q} = (2\pi, 2\pi) - 2\kN = (3.34, 3.34)$ for the N, which is a nesting vector connecting $\kN$ and $(2\pi, 2\pi) - \kN$.
For the HS, the behavior is in between.
The difference between AN and N stems from two factors.
First, $\im G(\mb{k}+\mb{q})$ (second row) has a plateau near $\mb{q} = (\pi, \pi)$ for the AN but is peaked at an incommensurate momentum for the N.
This happens as the AN is close to the van Hove singularity and has a flatter band dispersion than the N~\cite{Wu2017, Krien2020}.
Second, the antiferromagnetic fluctuations are short-ranged~\cite{Simkovic2024}, and the susceptibility decays slowly from the peak at $\mb{q} = (\pi, \pi)$.
Indeed, by fitting the susceptibility to a Lorentzian form $\chi_\Mch(\omega = 0, \mb{q}) = A / (\lvert \mb{q} - (\pi, \pi) \rvert^2 + \xi^{-2})$, we find the correlation length $\xi = 5.0$, much shorter than that for the weak-coupling PG at $U=2$ ($\xi > 20$)~\cite{Schaefer2021}.
The combination of these two factors enables incommensurate momentum transfer to play an important role in the self-energy, leading to the differentiation of the AN and the N.

Importantly, the above description is equally valid in the \lDGA.
Thus, the \lDGA shows the same hierarchy of AN, HS, and N self-energies as the p\DGA (\Fig{fig:Wu_self_energy}).
However, the \textit{absolute} scale of the momentum dependence in the self-energy is different between \lDGA and p\DGA due to the lack of enhancement of the Hedin vertex in the former [\Fig{fig:Wu_diagnostics}(b)].
Therefore, the N/AN differentiation from the transfer momentum diagnostics shown in \Fig{fig:Wu_diagnostics_in_q} is not sufficient for the strong-coupling PG; the enhanced Hedin vertex is crucial to have an actually insulating AN.

The PA result is qualitatively different.
The magnetic susceptibility in the PA is long-ranged and sharply peaked at $\mb{q} = (\pi, \pi)$.
Hence, only commensurate momentum transfer $\mb{q} = (\pi, \pi)$ gives a meaningful contribution to $\Sigma$.
The momentum dependence of $\Sigma$ is thus driven by nesting, and $\Sigma$ is enhanced also at the HS.
Figure~\ref{fig:Wu_PA_and_trilex}(b) confirms this: the PA $\Sigma$ fails to produce the hierarchy of AN, HS, and N, overestimating the insulating behavior at the HS.

\subsection{Additional results for the HM at PHS}

\begin{figure}[tb]
    \centering
    \includegraphics[width = 0.99\linewidth]{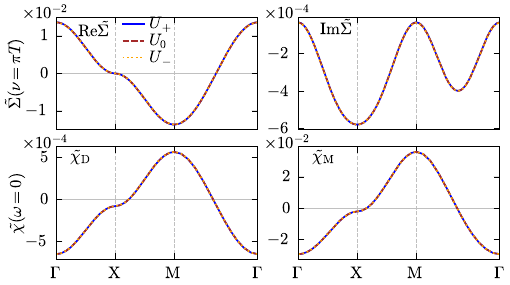}
    \caption{
    Self-energy and susceptibility as in Fig.~\ref{fig:highT}(c) computed with (fd-)p\DGA for three $U$ values near the vertex divergence ($U_0 / t = 8.356$): $U_+ = U_0 + 0.002 t$, $U_0$, and $U_- = U_0 - 0.002 t$.
    All curves lie on top of each other.
    }
    \label{fig:highT_various_U}
\end{figure}

\begin{figure}[tb]
    \centering
    \includegraphics[width = 0.99\linewidth]{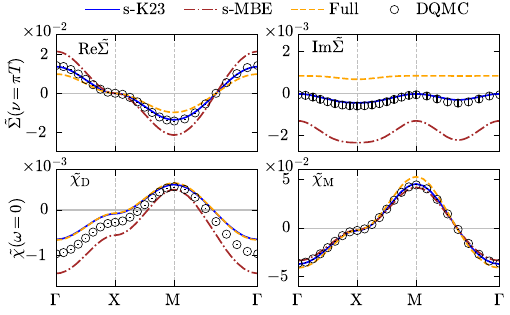}
    \caption{
    Self-energy and susceptibility as in Fig.~\ref{fig:highT}(c), with different approximations for the fermionic momentum dependence of the reducible vertices.
    s-K23: $s$-wave truncation for $K_2$, $K_{\bar{2}}$, $K_3$.
    s-MBE: $s$-wave truncation for the multiboson vertex, full momentum dependence for the Hedin vertex.
    Full: full momentum dependence with no approximation. 
    }
    \label{fig:highT_nonlocal}
\end{figure}

Figure~\ref{fig:highT_various_U} shows results analogous to \Fig{fig:highT}(c), for three $U$ values around the irreducible vertex divergence at $U_0$.
The results lie on top of each other, showing that fd-\DGA is agnostic with respect to irreducible vertex divergences.

In the main text, we used the $s$-wave approximation for the reducible vertex, dropping any fermionic momentum dependence.
We call this the s-K23 approximation, as the $K_2$, $K_{\bar{2}}$, and $K_3$ asymptotic classes are affected.
Here, we take a closer look at the effect of this approximation for the HM at PHS and a high temperature where the calculation is more tractable.
Following Refs.~\cite{Krien2021, Krien2022}, we adopt the single-boson-exchange decomposition of the vertex~\cite{Krien2019b} and apply the $s$-wave approximation only on the multi-boson exchange vertex (s-MBE).
We also perform a calculation with the full $\mb{q}$, $\mb{k}$, and $\mb{k'}$ dependence of reducible vertices.
Figure~\ref{fig:highT_nonlocal} shows that, while all calculations show a qualitatively similar trend, the $s$-K23 approximation, which involves the highest degree of approximation, yields the best agreement with the DQMC benchmark.
This implies an error cancelation between the underlying approximation of p\DGA (using single-site DMFT as the reference input for the fully 2PI vertex) and the $s$-wave approximation of the reducible vertex.

\subsection{Additional results for the doped HM}

\begin{figure}[tb]
    \centering
    \includegraphics[width = 0.99\linewidth]{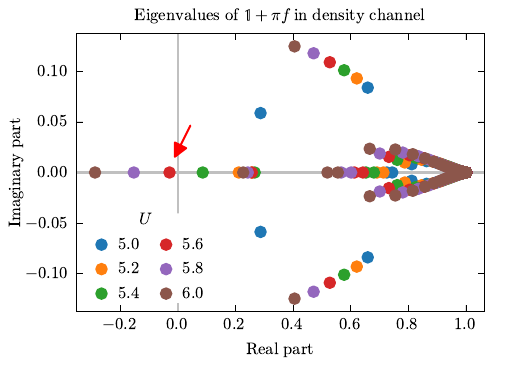}
    \caption{
    Eigenvalues of the matrix $X_{nn'} = \delta_{nn'} + \pi(i\nu_n) f(i\omega=0, i\nu_n, i\nu_{n'})$ in the density channel for the DMFT solution of the HM with parameters as in \Fig{fig:Wu_self_energy} ($U=5.6$) and for different values of $U$.
    Since the PHS is broken, the eigenvalues are either real or come in complex conjugate pairs~\cite{Essl2024}.
    At $U = 5.6$, one of the eigenvalues is very close to zero (marked by the red arrow), indicating that the system is close to a 2PI vertex divergence.
    }
    \label{fig:Wu_eigenvalues}
\end{figure}

\begin{figure}[tb]
    \centering
    \includegraphics[width = 0.99\linewidth]{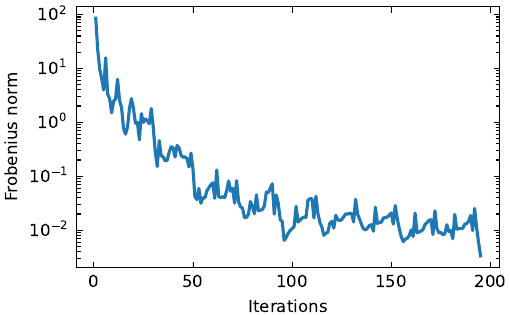}
    \caption{
    Convergence of the residuals of the fd-p\DGA calculation of the HM with parameters as in \Fig{fig:Wu_self_energy}.
    }
    \label{fig:Wu_convergence}
\end{figure}

\begin{figure}[tb]
\centering
\includegraphics[width = 0.99\linewidth]{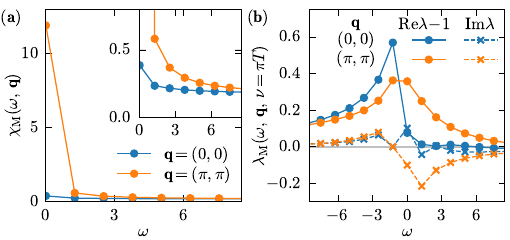}
\caption{
    The bosonic frequency dependence of (a) the magnetic susceptibility and (b) the magnetic Hedin vertex for the HM with parameters as in \Fig{fig:Wu_self_energy}.
    (See \Fig{fig:Wu_diagnostics_in_q} for their momentum dependence and the inset of \Fig{fig:Wu_diagnostics}(b) for the fermionic frequency dependence of the Hedin vertex. See also \Fig{fig:Wu_Hedin_2d}.)
}
\label{fig:Wu_Omega_dependence}
\end{figure}

\begin{figure}[tb]
\centering
\includegraphics[width = 0.99\linewidth]{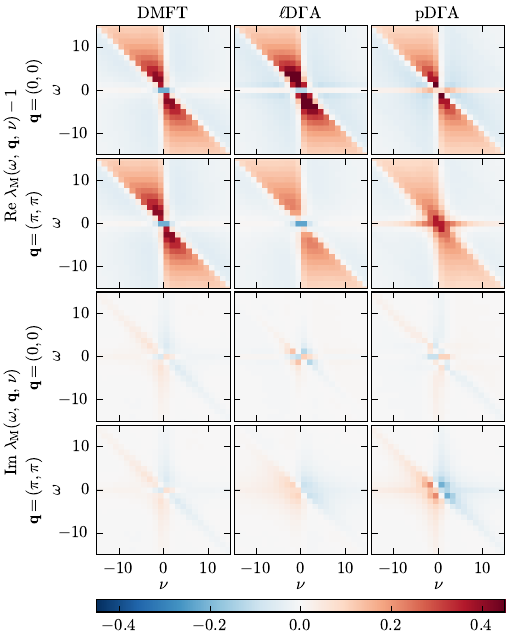}
\caption{
    The frequency dependence of magnetic Hedin vertices for the HM with parameters as in \Fig{fig:Wu_self_energy}, computed from DMFT, \lDGA, and p\DGA.
}
\label{fig:Wu_Hedin_2d}
\end{figure}

Figure~\ref{fig:Wu_eigenvalues} shows the eigenvalues of the matrix $\one + \pi f$ which enters the BSE for the DMFT irreducible vertex: $i = f (\one + \pi f)^{-1}$.
As this matrix is inverted in the BSE, a zero eigenvalue indicates a divergence of the irreducible vertex.
We find that $U=5.6$, the value we used in the main text to study PG, has an eigenvalue close to zero.
This shows its proximity to a 2PI vertex divergence and the importance of non-perturbative effects.

Figure~\ref{fig:Wu_convergence} shows the convergence of the fd-p\DGA fixed-point iteration using the restarted Anderson acceleration method~\cite{Anderson1965,Walker2011}.

It displays a Fermi arc with enhanced (suppressed) spectral weight near the node (antinode), which is a signature of the PG state.
Figure~\ref{fig:Wu_Omega_dependence} shows the bosonic frequency dependence of the susceptibility and Hedin vertex in the magnetic channel, and Fig.~\ref{fig:Wu_Hedin_2d} compares the two-dimensional frequency dependence of the magnetic Hedin vertex between DMFT, \lDGA, and p\DGA.

\clearpage

\end{document}